\documentclass[journal,twoside]{IEEEtran}

\usepackage{cite}

\ifCLASSINFOpdf
   \usepackage[pdftex]{graphicx}
  \DeclareGraphicsExtensions{.pdf,.jpeg,.png,.eps}
\else
  \usepackage[dvips]{graphicx}
  \usepackage{epsfig}
  \DeclareGraphicsExtensions{.eps,.pdf,.jpeg,.png}
\fi
\usepackage{amssymb}
\usepackage[cmex10]{amsmath}
\ifCLASSOPTIONcompsoc
  \usepackage[caption=false,font=normalsize,labelfont=sf,textfont=sf]{subfig}
\else
  \usepackage[caption=false,font=footnotesize]{subfig}
\fi
\usepackage{url} 
\usepackage{flushend} 
\usepackage{upgreek}

\begin{document}
%
\title{Development of High Precision Timing Counter Based on Plastic Scintillator with SiPM Readout}
%
%
%
\author{Paolo~W.~Cattaneo, 
        Matteo~De~Gerone,
        Flavio~Gatti,~\IEEEmembership{Member,~IEEE,}
        Miki~Nishimura, 
        Wataru~Ootani,
        Massimo~Rossella,
        and Yusuke~Uchiyama
\thanks{Manuscript received February 06, 2014; revised June 30, 2014; accepted August 09,2014.
This work was supported in part by MEXT KAKENHI Grant No.~22000004.}%
\thanks{P.~W.~Cattaneo and M.~Rossella are with the INFN of Pavia, Pavia I-27100, Italy.}
\thanks{M.~De Gerone  and F.~Gatti are with the INFN and University of Genova, Genova I-16146, Italy.}
\thanks{M.~Nishimura, W.~Ootani and Y.~Uchiyama are with the ICEPP, the University of Tokyo, Tokyo 113-0033, Japan (telephone: +81-3-3815-8384, e-mail: uchiyama@icepp.s.u-tokyo.ac.jp).}
}
%
%

\markboth{IEEE Transactions on nuclear science,~Vol.~, No.~, February~2014}%
{Cattaneo \MakeLowercase{\textit{et al.}}: Development of High Precision Timing Counter Based on Plastic Scintillator with SiPM Readout}
%

\IEEEpubid{0000--0000/00\$00.00~\copyright~2014 IEEE}

\maketitle

\newcommand*{\megsign}        {\mu^+ \to \mathrm{e}^+ \gamma}

\begin{abstract}
\boldmath
High-time-resolution counters based on plastic scintillator with silicon photomultiplier (SiPM) readout have been developed for applications to high energy physics experiments for which relatively large-sized counters are required. 
We have studied counter sizes up to $120\times40\times5~\mathrm{mm}^3$ with series connection of multiple SiPMs to increase the sensitive area and thus achieve better time resolution.
A readout scheme with analog shaping and digital waveform analysis is optimized to achieve the highest time resolution. 
The timing performance is measured using electrons from a $^{90}$Sr radioactive source, comparing 
different scintillators, counter dimensions, and types of near-ultraviolet sensitive SiPMs.
As a result, a resolution of $\sigma =42 \pm 2$~ps  at 1~MeV energy deposition is obtained for counter size $60\times 30 \times 5~\mathrm{mm^3}$ with three SiPMs ($3\times3~\mathrm{mm^2}$ each) at each end of the scintillator.
The time resolution improves with the number of photons detected by the SiPMs. The SiPMs from Hamamatsu Photonics give the best time resolution because of their high photon detection efficiency in the near-ultraviolet region.
Further improvement is possible by increasing the number of SiPMs attached to the scintillator.
\end{abstract}

\begin{IEEEkeywords}
scintillation counters, time resolution, silicon photomultiplier (SiPM), Multi-Pixel Photon Counter (MPPC).
\end{IEEEkeywords}

%

\section{Introduction}
%
%
%
%
\IEEEPARstart{V}{ery} precise time measurement is one of the important ingredients in a wide range of physics experiments. 
Scintillation counters with photomultiplier tube (PMT) readout have been widely used for this purpose.
However, silicon photomultipliers (SiPMs) can be a good replacement for PMTs as photo-sensors for use with  scintillation counters because of their high photon detection efficiency (PDE) and their high single photon time resolution (SPTR).
Furthermore, other SiPM features such as compactness, insensitivity to magnetic field,
and low cost allow major improvements to detector design. 

Our interest is focused on developing a timing counter system to measure the time of positrons of $\sim\!50$~MeV/c with an ultimate time resolution of $\lesssim\!30$~ps\footnote{Resolutions are always quoted as RMS deviation.} in the MEG~II experiment \cite{megII,megup}.
The concept of the new detector is to segment
the timing counter system into several hundred small scintillation counters, each of which is readout by several SiPMs. Each particle\rq{s} time is measured by multiple counters, significantly improving the resolution with respect to that of a single counter \cite{megup,ootani_nima}. 
This work focuses on the study and optimization of the time resolution of a single counter.

\IEEEpubidadjcol

Excellent time resolution, $\sigma= 18/\sqrt{E/(1\,\mathrm{MeV})}~\mathrm{ps\,}$, 
was reported for a counter based on plastic scintillator with SiPM readout in \cite{muSR}. Their counters, with dimensions $3\times3\times2~\mathrm{mm^3}$, are too small for our application.
In this work, we investigate the achievable time resolution for larger scintillation counters, with dimensions from $60\times30\times5$ to $120\times40\times5~\mathrm{mm}^3$.

In considering larger detector sizes, the main drawback of SiPM-based readout is their small active area. 
Devices with an active area of $3\times3$~$\mathrm{mm^2}$ are now widely available, while development of larger SiPMs, including SiPM arrays, is progressing.
Parallel connection among the sensors in an array, which is equivalent to a single large SiPM from the circuit viewpoint, is often adopted to sum up the signals.
Performance issues for such large-sized SiPMs are an increase in the parallel and series noise, non-uniform spatial response, and increase in the signal rise time and width, all of which may worsen the time measurement. 
Most of these issues originate from the parallel connection. 
Though individual readout of each sensor chip can be a solution, this scheme results in an increase in the number of readout channels and is not an appropriate solution for our application.
In this work, we adopt a different scheme for the SiPM connection -- connecting $3\times3$~$\mathrm{mm^2}$-sized SiPMs in series -- to be nearly insensitive to these issues on the time measurement.

SiPMs generally have good SPTR (e.g. \cite{buzhan_nima,gundacker_nima}). 
In addition, a high PDE in the near-ultraviolet (NUV) region is required for high time resolution in fast plastic scintillator readout. 
Recently, several manufacturers have provided NUV-sensitive SiPMs based on \lq{}p-on-n\rq{} diode structures.
We therefore tested a number of such NUV-sensitive SiPMs: AdvanSiD NUV-type \cite{fbk}, KETEK SiPM \cite{ketek}, SensL B-series with fast output \cite{senslB} and Hamamatsu Photonics (HPK) MPPCs. The latter includes recently developed types with technologies for after-pulse and/or cross-talk suppression \cite{hamamatsu,hamamatsu_nima}.

\section{Experimental details}
\subsection{Setup for time resolution measurements}

\begin{figure}[!t]
\centering
\includegraphics[width=3.5in]{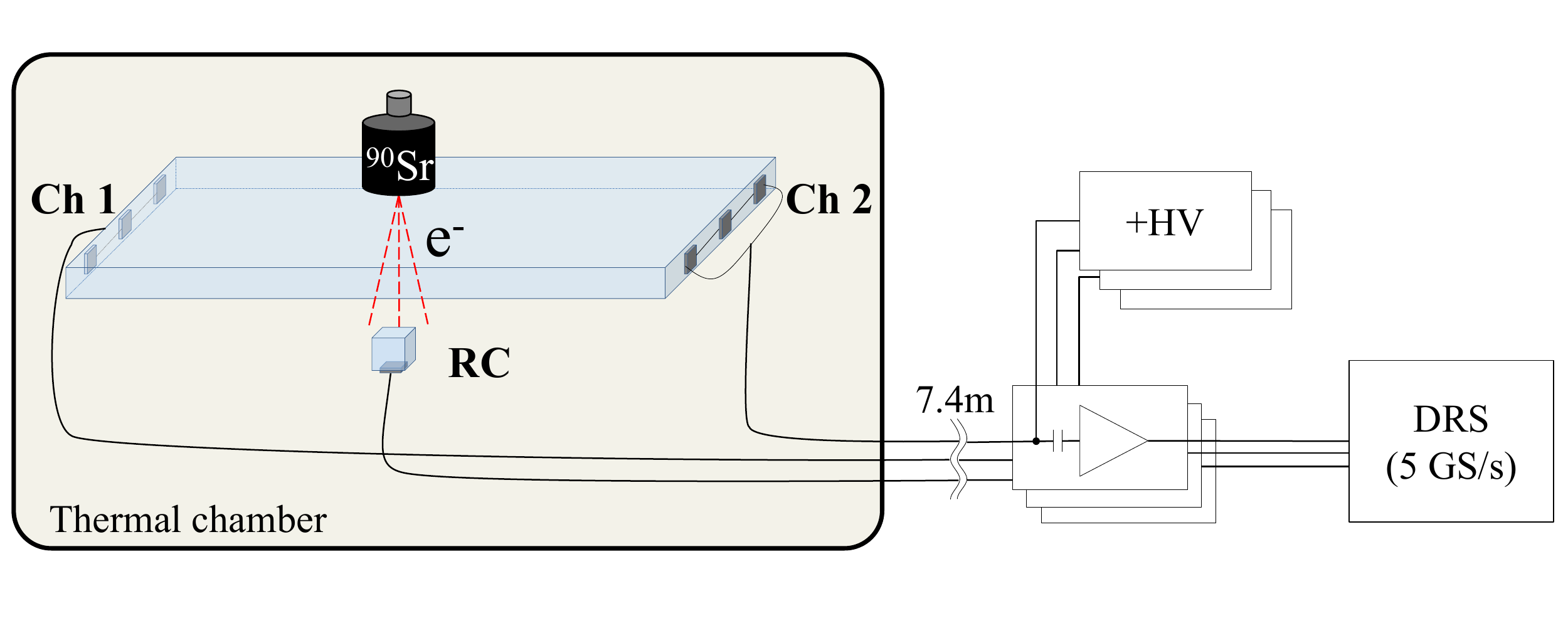}
\caption{Test setup for measurements of the counter time resolution. RC denotes the reference counter. See the text for details.}
\label{fig:setup}
\end{figure}

The setup for measuring the time resolution of scintillation counters is schematically shown in Fig~\ref{fig:setup}.
A test counter is composed of a scintillator plate and six SiPMs.
The counter dimensions are defined by the length ($L$), width ($W$) and thickness ($T$) of the scintillator plate, and always written as $L\times W\times T$ in this paper.
Three SiPMs are optically coupled to each $W\times T$ plane
of the scintillator with optical grease (OKEN6262A).
The signals from the three SiPMs are summed and readout on a channel as described in detail later.

The counter is irradiated by electrons from a $^{90}$Sr source ($E_\mathrm{e}<2.28$~MeV), and the impact point is selected by means of a small ($5\times5\times5~\mathrm{mm}^3$) reference counter placed behind the test counter.  The reference counter is made from BC422, wrapped in Teflon tape and readout by an HPK SiPM S10362-33-050C. The time measured by this reference counter 
is used as a time reference.
The mean energy deposited in the test counter (5~mm thick) is evaluated to be 0.95~MeV by a Monte Carlo simulation for events selected by requiring more than 0.5~MeV energy to be deposited in the reference counter.
These counters are put in a thermal chamber at a constant temperature ($23^\circ$C for the standard measurement).

\begin{figure}[!t]
\centering
\includegraphics[width=3.5in]{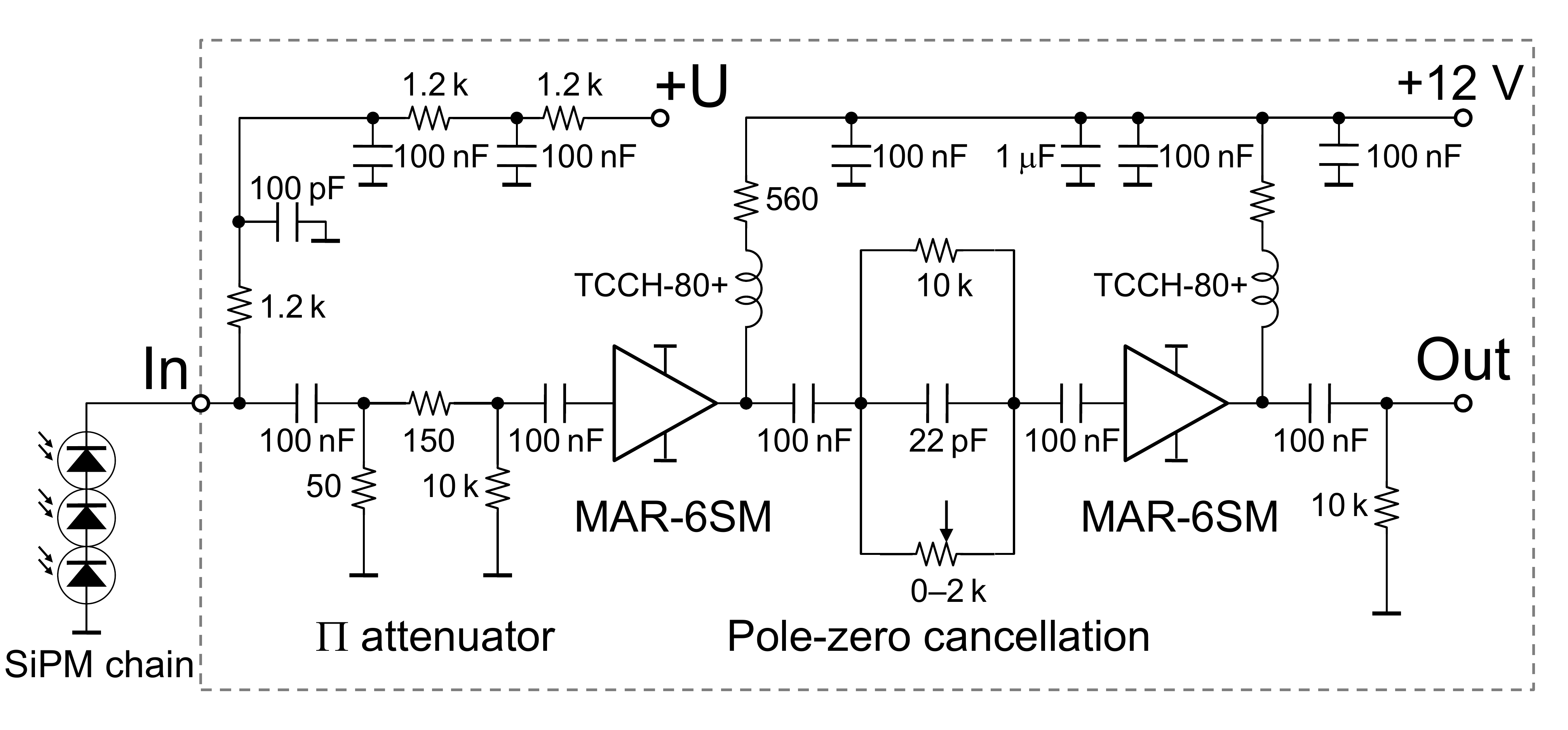}
\caption{Circuit diagram of the amplifier board with equivalent input impedance of $40~\Omega$, bandwidth of $800$~MHz (3~dB limit), and overall transimpedance gain of $970~\Omega$.}
\label{fig:amplifier}
\end{figure}

\begin{figure}[!t]
\centering
\includegraphics[width=3.in]{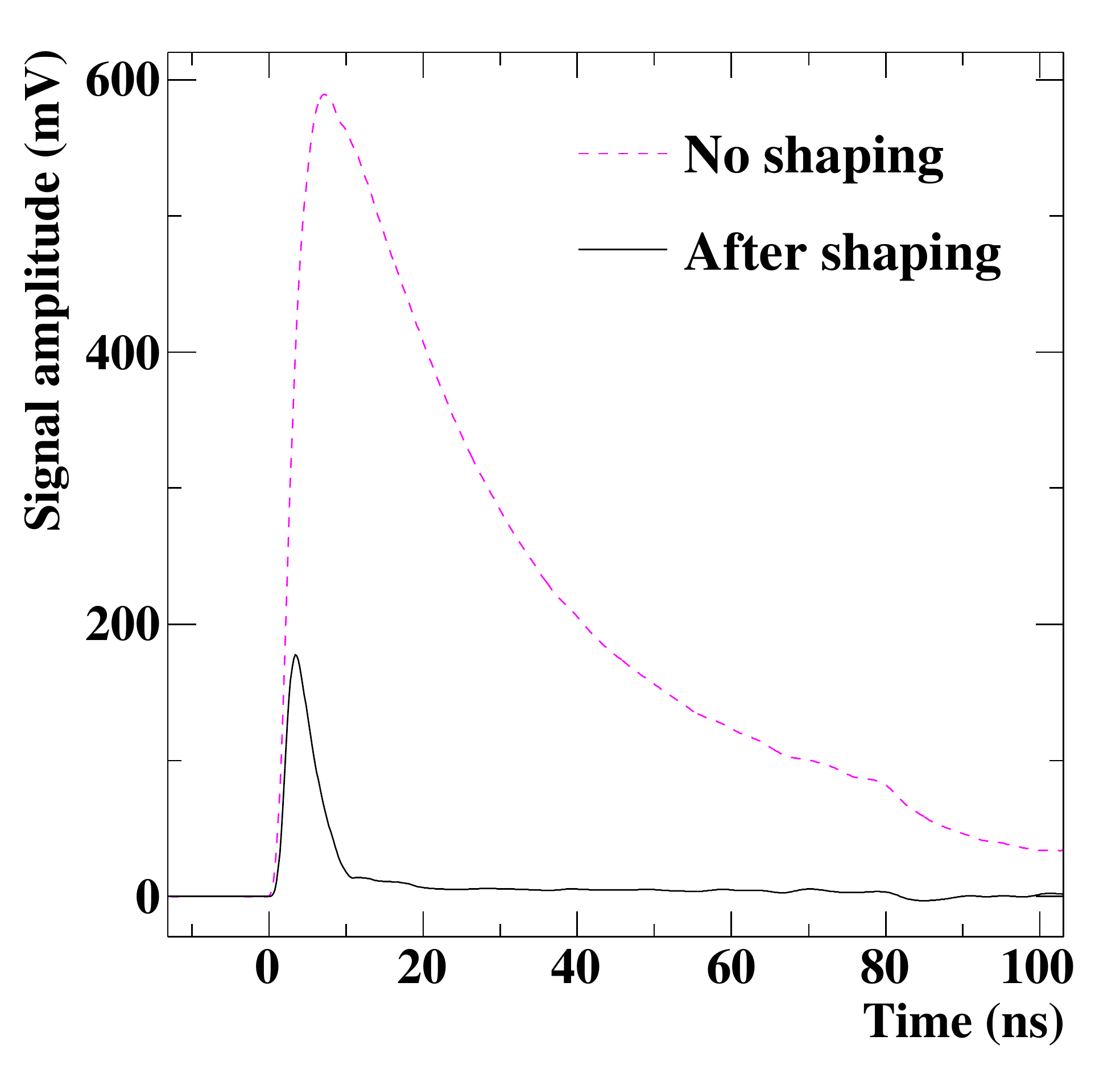}
\caption{Comparison of average pulse shapes of scintillation signal (BC422) with HPK SiPM readout (S10931-050P with 3 in series and 2.0~V over-voltage) with (solid line) and without (dashed line) the analog shaping. These pulse shapes are obtained by averaging the signal waveform data over several thousand events.}
\label{fig:shapedWf}
\end{figure}

The signal from the SiPM chain at each end of the counter is transmitted on a 7.4-m long coaxial cable (standard RG174 type) to an amplifier and then readout by the fast sampling waveform digitizer DRS4 \cite{drs}, mounted on the DRS4 evaluation board V4 with an analog bandwidth of 750 MHz \cite{drsboard}.
This electronics chain, with SiPMs and amplifiers separated by a long cable, is convenient for many applications because of space and other environmental limitations at the counter.
This setup simulates that expected in the MEG II experiment. 
The amplifier\footnote{The circuit was designed at the Paul Scherrer Institut by U.~Greuter}, whose circuit schematic is shown in Fig.~\ref{fig:amplifier}, is based on a two-stage voltage amplifier with pulse shaping. 
The analog pulse shaping is optimized for time measurement and uses a pole-zero cancellation circuit to select the fast, leading-edge part of the signal and to restore quickly a stable baseline, as shown in Fig.~\ref{fig:shapedWf}.

The biasing scheme for the SiPMs is also shown in Fig.~\ref{fig:amplifier}.
A positive bias voltage from the amplifier board is supplied to the SiPM chain through the signal line.
In this bias and readout scheme, the polarity of the SiPM signal is negative.
To match the DRS4 dynamic range, the signal is inverted at its input by a transformer (ORTEC IT100, bandwidth of 440~MHz). 

The signal time is measured by analyzing the waveform as described in Sec.~\ref{drswaveform}.
The electron impact time is computed by the average of the signal times measured at the two ends of the test counter, $t_\mathrm{counter} = (t_1 + t_2)/2$.


\subsection{Setup for SiPM characterization}
In a preliminary study, we investigated the single sensor performance of each type of SiPM, not attached to the scintillator. 
In this study, the pulse shaper on the amplifier board was bypassed in order to get a higher signal-to-noise ratio in the charge measurement of small signals.

Dark signal data were taken by random triggering.
A dark count rate is calculated from the probability of observing zero fired pixels
$P(N_\mathrm{fired}=0)$ in a fixed time window. 
Assuming Poisson statistics, we calculate the average number of dark counts in the time window as
$\lambda_\mathrm{dark}=-\ln{P(N_\mathrm{fired}=0)}$.

A cross-talk probability is calculated from the ratio $P(N_\mathrm{fired}\geq2)/P(N_\mathrm{fired}\geq1)$ after correcting for the accidental coincidence of dark pulses.

For a relative comparison of the SiPM PDEs for NUV light, an ultraviolet LED (Toyoda Gosei E1S19-0P0A7), with wavelength (370--410~nm) approximately matching that of fast plastic scintillators, is installed in the thermal chamber. 
The SiPM signal was taken in synchronization with the timing of the LED pulsing. 
The LED intensity was adjusted so that the resultant average number of fired pixels for the HPK SiPM (S10362-33-050C) ranged roughly between 0.5 and 1.0 and fixed over all the measurements.
We calculate the relative PDE from $P(N_\mathrm{fired}=0)$ in the LED data with a correction for the accidental coincidence of dark pulses in accordance with the Poisson statistics.

\subsection{SiPM under test}

 Table~\ref{tab:sample} lists the NUV SiPMs tested in this work.
All of them are based on the p-on-n diode structure and have the same active area, $3\times3$~$\mathrm{mm^2}$.
\begin{table*}[!t]
\renewcommand{\arraystretch}{1.3}
\caption{List of SiPM test samples.}
\label{tab:sample}
   \newcommand{\m}{\hphantom{$-$}}
\centering
  \begin{minipage}{0.7\linewidth}
   \renewcommand{\thefootnote}{\alph{footnote})}	
   \renewcommand{\thempfootnote}{\alph{mpfootnote})}	
   \centering
\begin{tabular}{@{}llll}
\hline
\textbf{Manufacturer}\m  & \textbf{Model number} & \multicolumn{2}{l}{\m \textbf{Type}\footnotemark[1]}\\
\hline \hline
HPK & S10362-33-050C & \m Conventional (Old ) MPPC & Ceramic package\\
	& S10931-050P & & Surface mount type (SMT) package\\ \cline{2-4}
	& S12572-050C(X)\footnotemark[2] & \m Standard-type (New) MPPC & Metal quench resistor \\
	& S12572-025C(X)\footnotemark[2] & & 25 $\upmu$m pixel \\ \cline{2-4}
	& S12652-050C(X)\footnotemark[2] & \m Trench-type MPPC & Metal quench resistor \\
	& 3X3MM50UMLCT-B\footnotemark[2] & & Improved fill factor\\
\hline
AdvanSiD & ASD-NUV3S-P-50\footnotemark[2] & \m NUV type &SMT package  \\
\hline
KETEK & PM3350 prototype-A\footnotemark[2] & \m Trench type & --\\
\hline
SensL & MicroFB-30050-SMT & \m B-series & With fast output. SMT package\\
\hline
\end{tabular}
  \vspace{-0.2cm}
     \footnotetext[1]{Sensor size of all the samples is $3\times3~\mathrm{mm}^3$. Pixel pitch is $50~\upmu\mathrm{m}$ unless specified.}
     \footnotetext[2]{Not a commercial product. Under development.}
\end{minipage}
\end{table*}

\subsection{SiPM connection}

We use a series connection to sum signals from the three SiPMs attached to each end of the scintillator. 
Series connection of avalanche photodiodes (APDs) was proposed in \cite{CMS} and 
 first applied to SiPMs in \cite{sipmSeries}.

\begin{figure}[!t]
\centering
\subfloat[Individual SiPM]{\includegraphics[width=3.in]{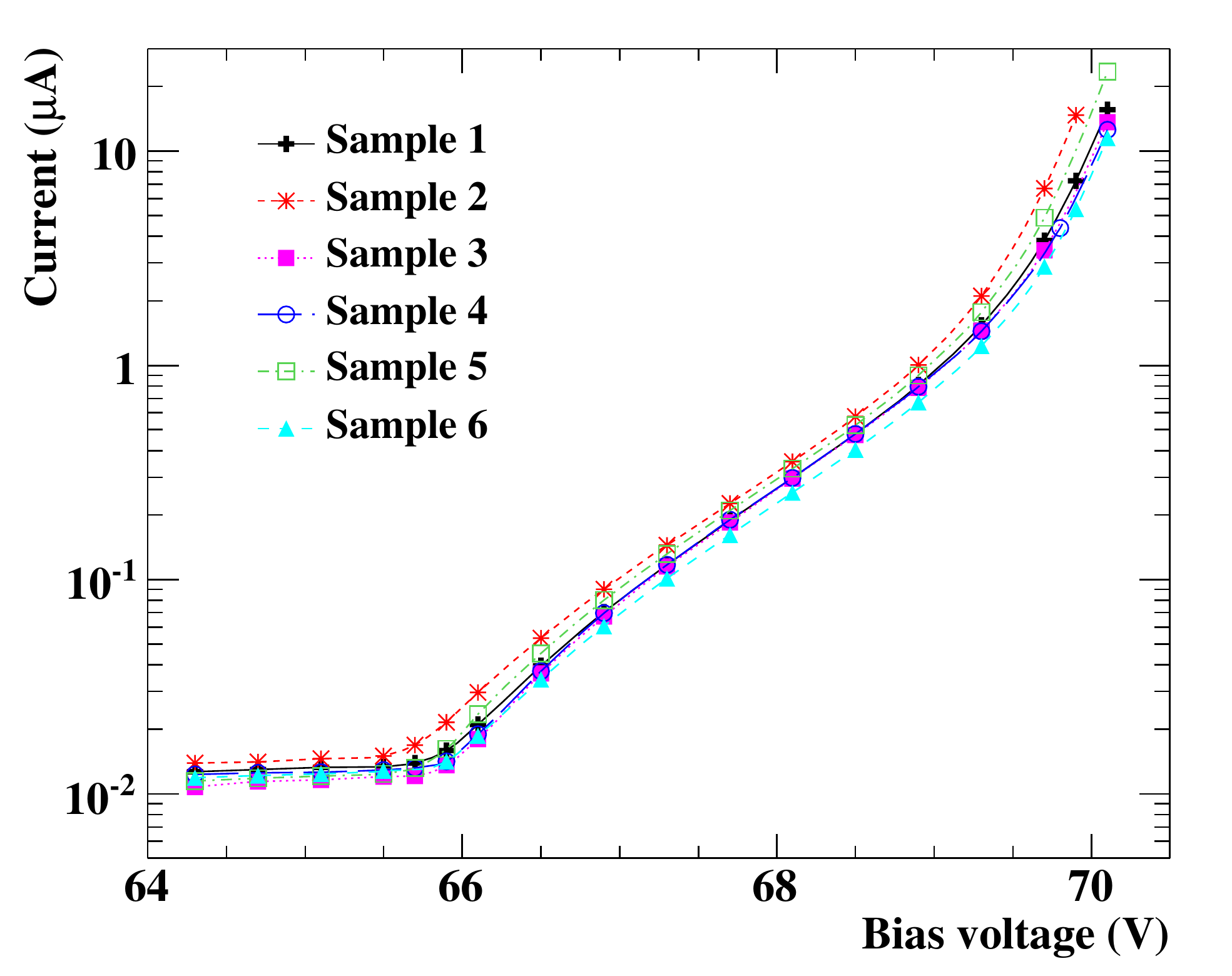}%
\label{fig:ivSingle}}
\hfil
\subfloat[3-SiPM series connection]{\includegraphics[width=3.in]{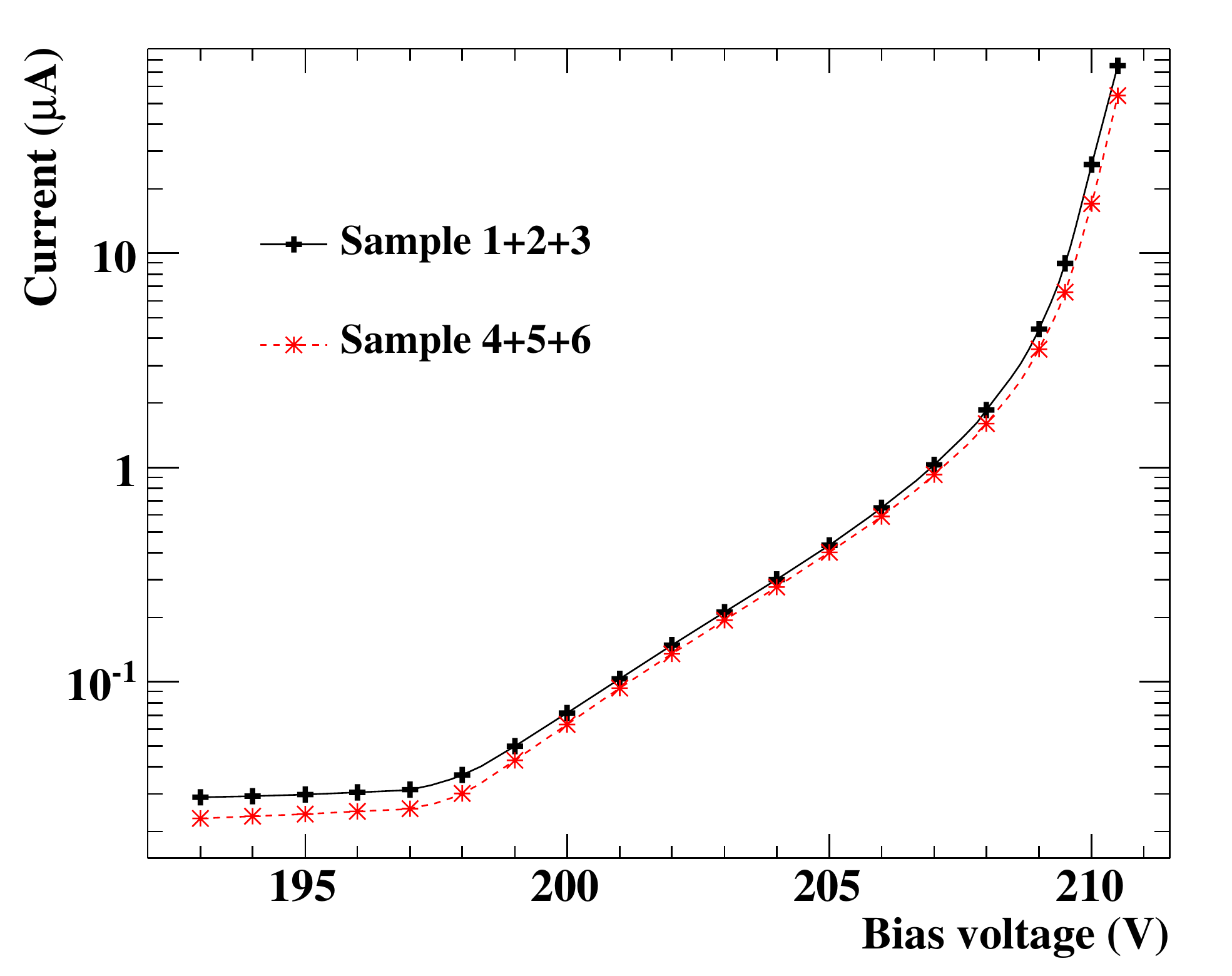}%
\label{fig:ivSeries}}
\caption{I-V characteristic curves of the new-type HPK SiPMs (S12572-050C(X)).}
\label{fig:iv}
\end{figure}

One of the advantages of the series connection compared with the more conventional parallel connection is the automatic adjustment of over-voltage among the three SiPMs, even if the individual breakdown voltages are different. Fig.~\ref{fig:ivSingle} shows examples of I-V characteristics of individual SiPMs. 
While the individual breakdown voltages differ by a few hundred millivolts, the shapes of the I-V curves are quite similar. 
When SiPMs are connected in series, the voltage applied to each SiPM is determined by the common leakage current. Then, the difference in breakdown voltages is absorbed, and the over-voltages are approximately aligned.
Fig.~\ref{fig:ivSeries} shows I-V characteristic curves for two examples of three SiPMs operated in a series configuration. 

\begin{figure}[!t]
\centering
\includegraphics[width=3.in]{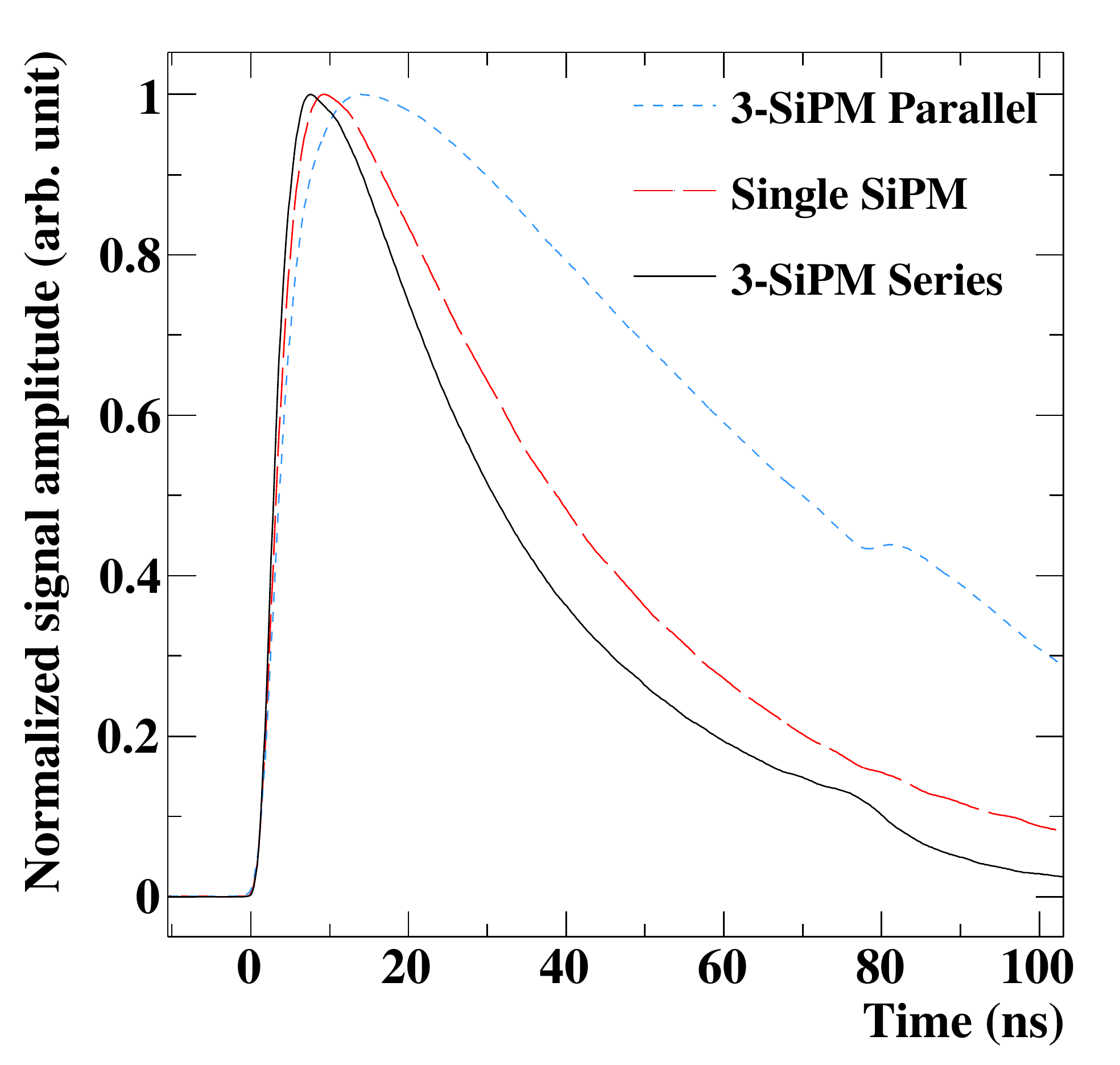}
\caption{Pulse shape comparison for different SiPM connections. Scintillation signals of BC422 with HPK SiPM (S10362-33-050C) readout without the analog pulse shaping are shown. These pulse shapes are made by averaging a few thousand events and normalized by their amplitudes. Bumps at around 80 ns are due to a reflection in the readout circuit.}
\label{fig:connectionWf}
\end{figure}

Another advantage of series connection is that the resultant pulse shape becomes narrower than that from a single SiPM, as shown in Fig.~\ref{fig:connectionWf}.
This is in contrast to the case of parallel connection, where the signal becomes wider. 
This is due to the reduction of the total capacitance of the series circuit consisting of the junction capacitances of the reverse-biased diodes and the associated stray capacitances.
The fast rise time is of particular importance for optimizing the time resolution. 
A disadvantage is that the signal size is reduced to one third of that of a single SiPM. 
Nevertheless, the photon-counting capability is retained in the 3-SiPM series connection.

The SensL sensors have a third pin (fast output) in addition to the anode and cathode pins \cite{sensl,pavlov2013silicon} that allows the extraction of a signal that is very fast compared with that of the signal from the anode--cathode line. 
Using this fast output is expected to have a positive impact on time measurement. 

\begin{figure}[!t]
\centering
\includegraphics[width=3.5in]{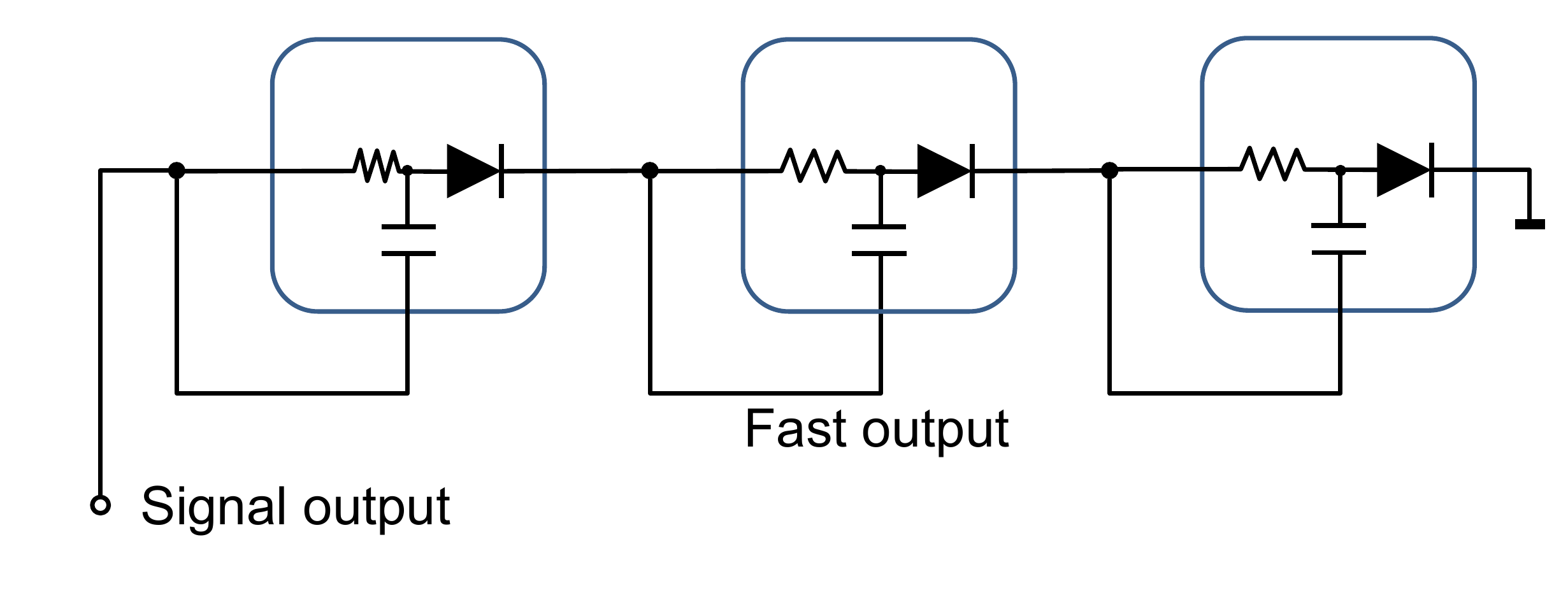}
\caption{Readout scheme of the SensL SiPMs. The output signal is transmitted to the amplifier where a negative bias voltage is also supplied.}
\label{fig:sensl}
\end{figure}

Fig.~\ref{fig:sensl} shows the readout scheme tested for the SensL SiPM.
The fast output is combined with the standard output in a single line; in this configuration a fast and uniform output of the sum signal can be obtained \cite{pavlovScheme}.
In this scheme, a negative bias is applied on the anode through the signal line.

\subsection{Scintillator}
Scintillator materials have been carefully investigated from the viewpoint of light yield, rise time, decay time, and emission spectrum.
We tested four types of ultra-fast plastic scintillators from Saint-Gobain Crystals: BC418, BC420, BC422, and BC422Q (0.5\% benzophenone as a quenching agent). 

A study to optimize the scintillator wrapping was carried out in our previous work \cite{ootani_nima}, and a specular reflector (3M radiant mirror film) was found to give the best resolution. However, no wrapping was adopted for this comparative study.

\subsection{Waveform analysis}
\label{drswaveform}
The signal time is measured by analyzing the waveform with a digital-constant-fraction method, an algorithm to pick off the pulse time at which the signal reaches a pre-determined fraction of the full amplitude \cite{dcf}.
\begin{figure}[!t]
\centering
\includegraphics[width=3.4in]{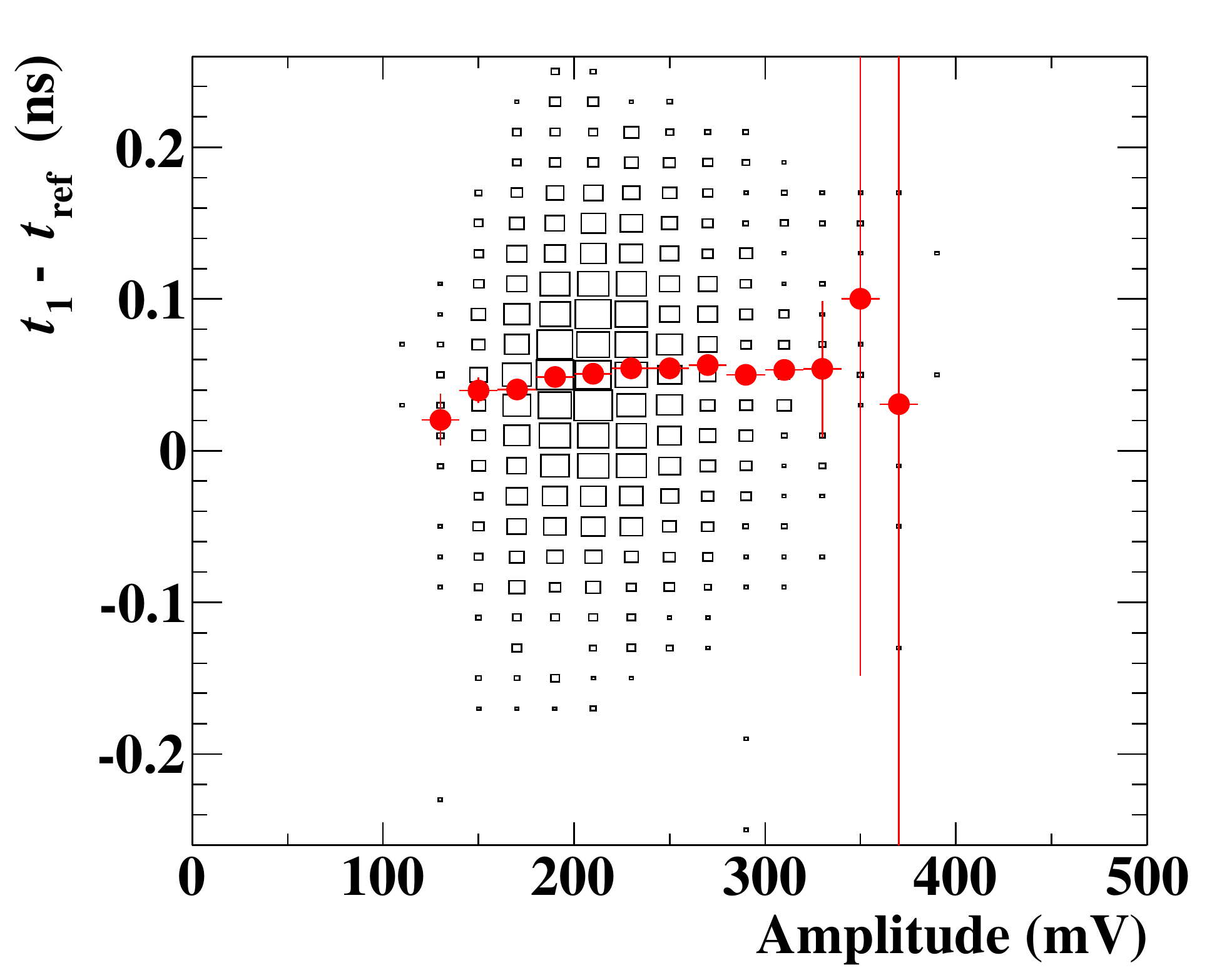}
\caption{Signal time dependence on the pulse amplitude by the digital-constant-fraction method ($60\times 30\times 5~\mathrm{mm^3}$ counter with S10362-33-050C, fraction at 6\%).}
\label{fig:cf}
\end{figure}
The stability of the measured signal time over different signal amplitudes is shown in Fig.~\ref{fig:cf}. 
The data show the nearly amplitude-independent timing that is achieved using this method in our signal amplitude range. 
The optimal value of the fraction ranges from 3\% to 6\% depending on SiPM type (for different rise times and signal-to-noise ratios). 

We also tested measuring the signal time as the time at which the pulse crosses a fixed threshold. 
The threshold values were scanned and optimized for each configuration.
The best resolutions are comparable to that by the digital-constant-fraction method when a proper time-walk correction is applied. 
However, the resolutions are sensitive to the threshold and the time-walk correction coefficients. Over the variable configurations, the digital-constant-fraction method gives more stable results, and thus, we adopt it as our default method.

In these measurements, the waveform data were acquired at $5~\mathrm{GS/s}$. We observed that the comparable resolution can be achieved at sampling speeds down to $2~\mathrm{GS/s}$ by applying a cubic interpolation among the samples in the digital-constant-fraction method.

\section{Results}

\subsection{SiPM characteristics}

\subsubsection{Noise measurements}
\begin{figure}[!t]
\centering
\subfloat[Dark count rate]{\includegraphics[width=3.5in]{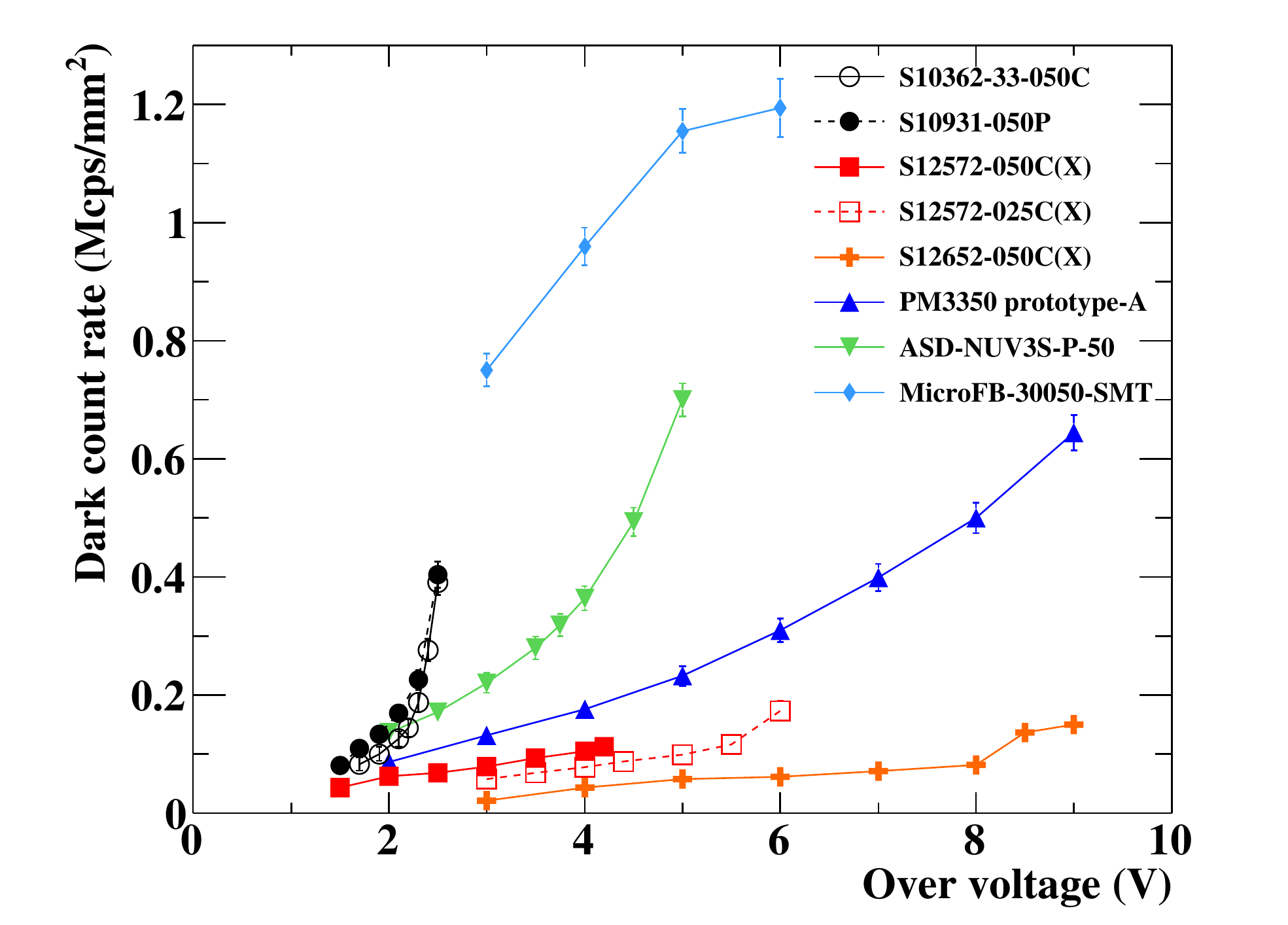}%
\label{fig:dark}}
\hfil
\subfloat[Cross-talk probability]{\includegraphics[width=3.5in]{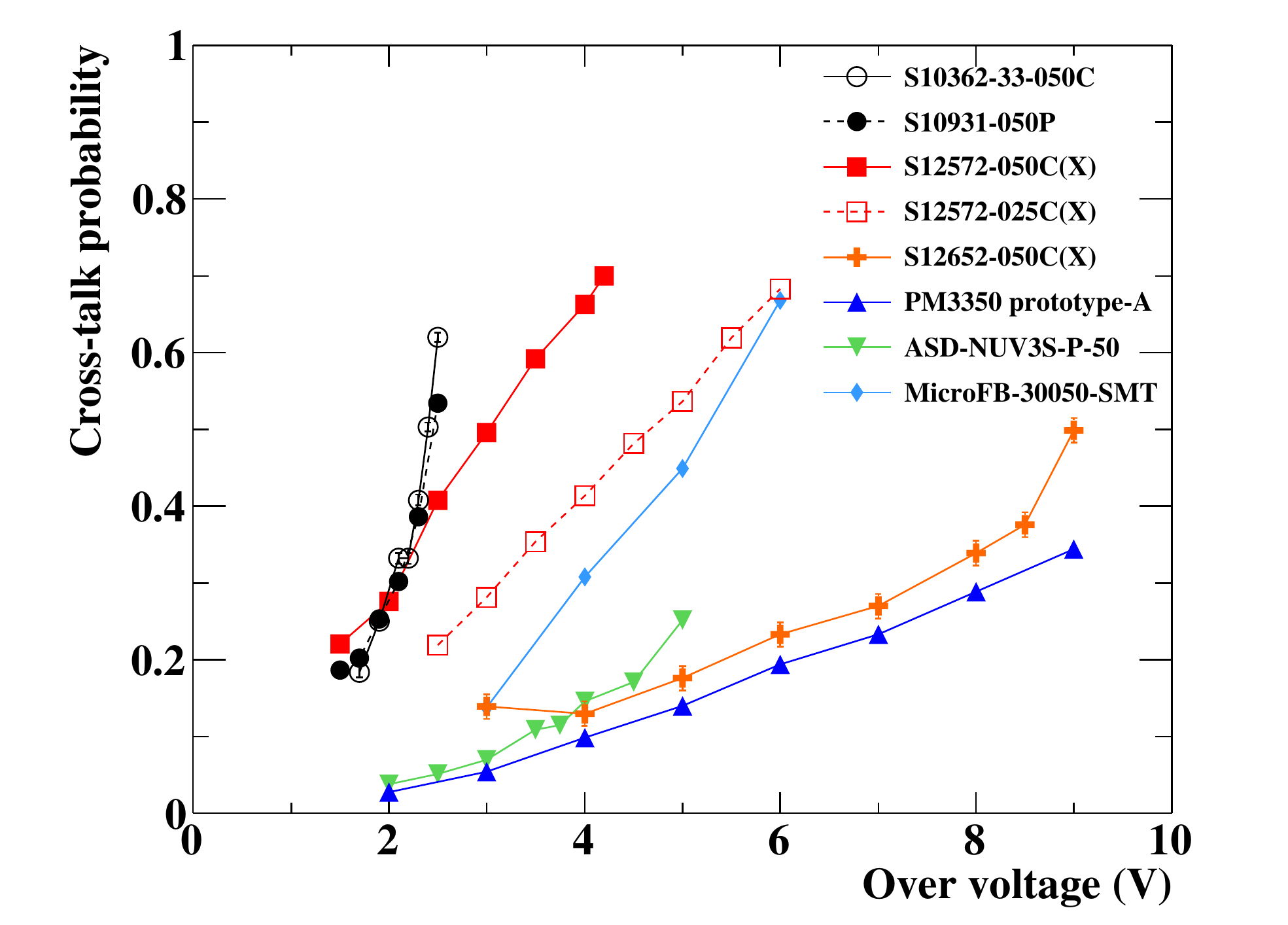}%
\label{fig:crosstalk}}
\caption{Results of noise measurements.}
\label{fig:noise}
\end{figure}

The measurements of the dark count rates are shown in Fig.~\ref{fig:dark} as a function of over-voltage.
Note that this measurement includes the effect of after-pulsing.
It is well known that the after-pulsing probability increases with the over-voltage, and therefore, the dark count rates show quadratic or higher growth with overvoltage.
However, the new-type HPK SiPMs show low dark count rates up to high over-voltages because of the suppression of after-pulsing.
\begin{figure}[!t]
\centering
\includegraphics[width=3.4in]{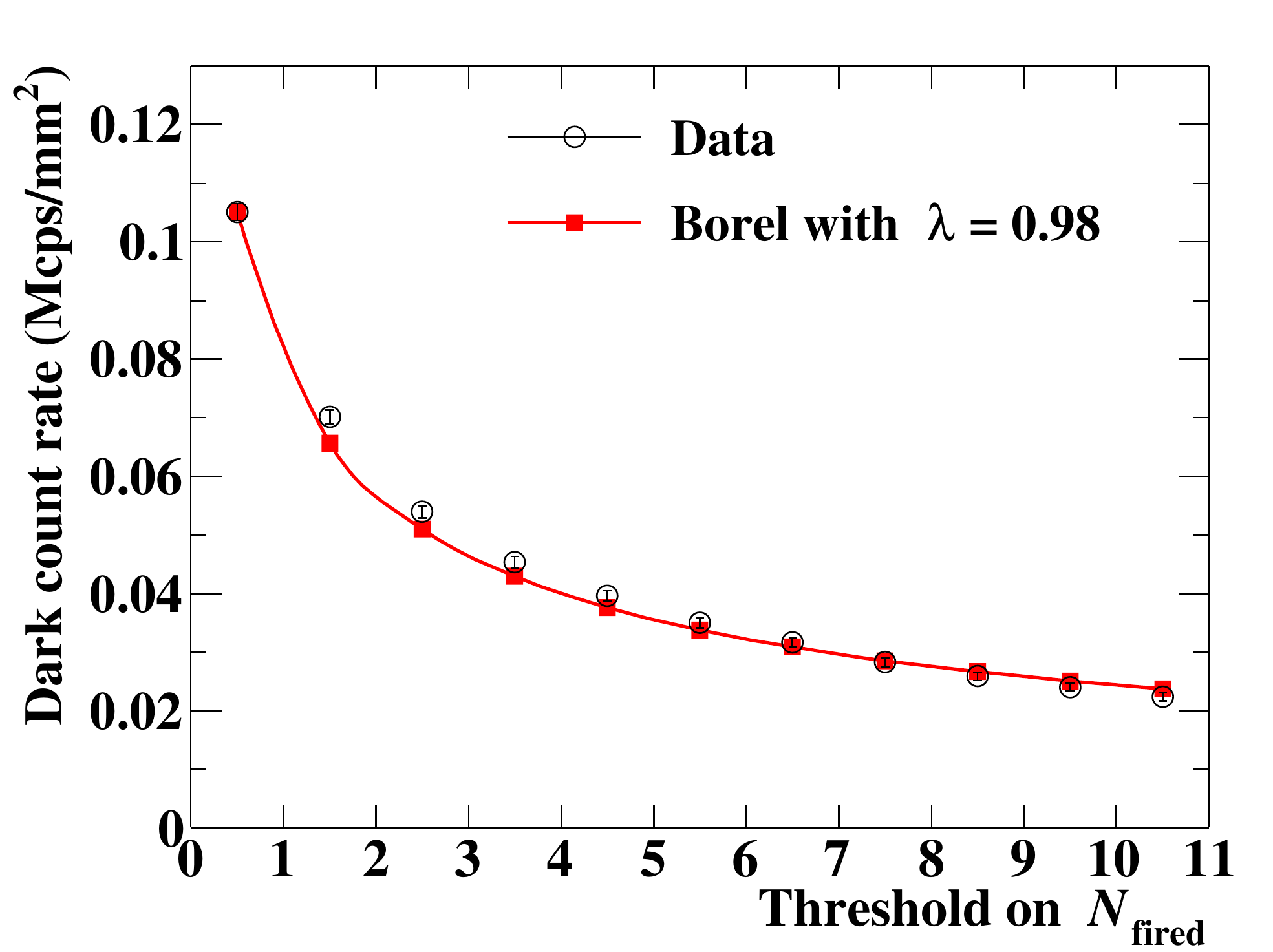}
\caption{Dependence of dark count rate on the threshold on the number of fired pixels, measured with S12572-050C(X) at an over-voltage of 4~V. The curve with square markers shows the analytical model of the cross-talk process with the branching Poisson process (cumulative distribution of the Borel distribution with the single generation parameter $\lambda=0.98$) \cite{crosstalkModel}.}
\label{fig:borel}
\end{figure}

Fig.~\ref{fig:crosstalk} shows the result of the cross-talk measurement.
The cross-talk probabilities increase almost linearly with the over-voltage.
Fig.~\ref{fig:borel} shows the dark count rate dependence on the threshold level on the number of fired pixels.
The probability distribution of the cross-talk process agrees fairly well with the branching Poisson process description \cite{crosstalkModel}.
The cross-talk probability of the standard-type SiPMs, namely SiPMs without a trench structure, reaches up to $\sim\!70$\% at which point the cross-talk process diverges. 
Such a high probability for the cross-talk process would generally become a problem for charge (energy) measurement due to the large excess noise factor. 
At the same time, it has an impact on time measurement as discussed in Sec.~\ref{discussiontr}.
On the other hand, the cross-talk probability for the trench-type SiPMs stays low up to over-voltages as high as 9~V.

\subsubsection{PDE measurements}

\begin{figure}[!t]
\centering
\vspace{-1mm}
\includegraphics[width=3.5in]{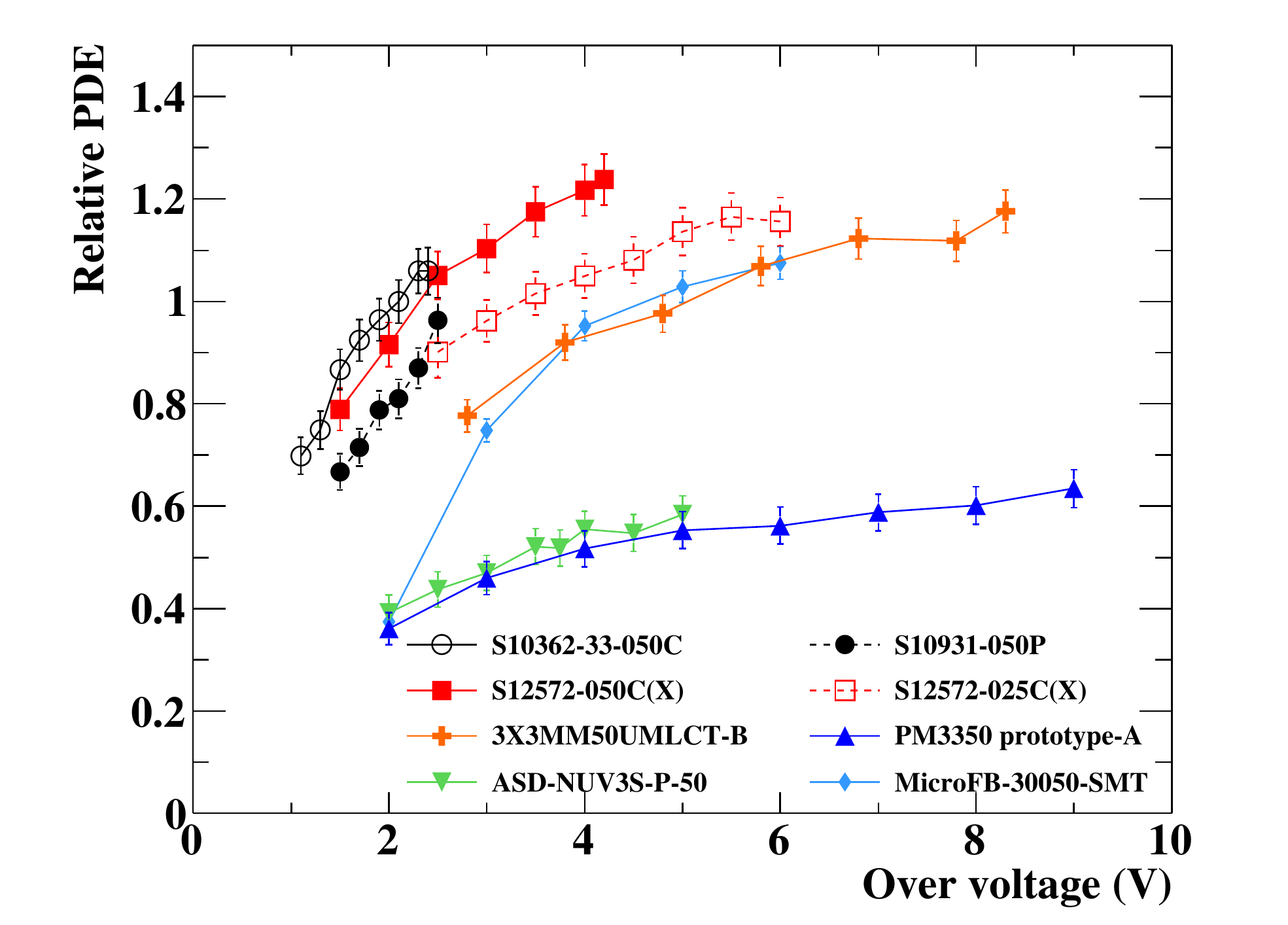}
\vspace{-5mm}
\caption{Results of relative PDE measurements in the NUV region using a UV-LED.}
\label{fig:pde}
\end{figure}

Fig.~\ref{fig:pde} summarizes the results of PDE measurements.
The systematic uncertainty from the LED instability and the positioning of the SiPMs relative to the LED is estimated to be 3\% by repeating the measurements.
The HPK and SensL SiPMs show higher PDEs than the others in the NUV region. The PDEs of the new-type HPK SiPMs are further improved in the extended over-voltage range.
The 25-$\upmu$m pixel SiPMs and the trench-type SiPMs from HPK possess comparable PDEs due to their improved fill factors. 

\subsection{Time resolution}
The time resolution of the test counter is evaluated from the distribution of difference between the time measured by the test counter and that by the reference counter ($\Delta t = t_\mathrm{counter} - t_\mathrm{ref}$).
The contribution of the time resolution of the reference counter is subtracted as $\sigma = \sqrt{\sigma_{\Delta t}^2 - \sigma_\mathrm{ref}^2}$.

\begin{figure}[!t]
\centering
\includegraphics[width=3.4in]{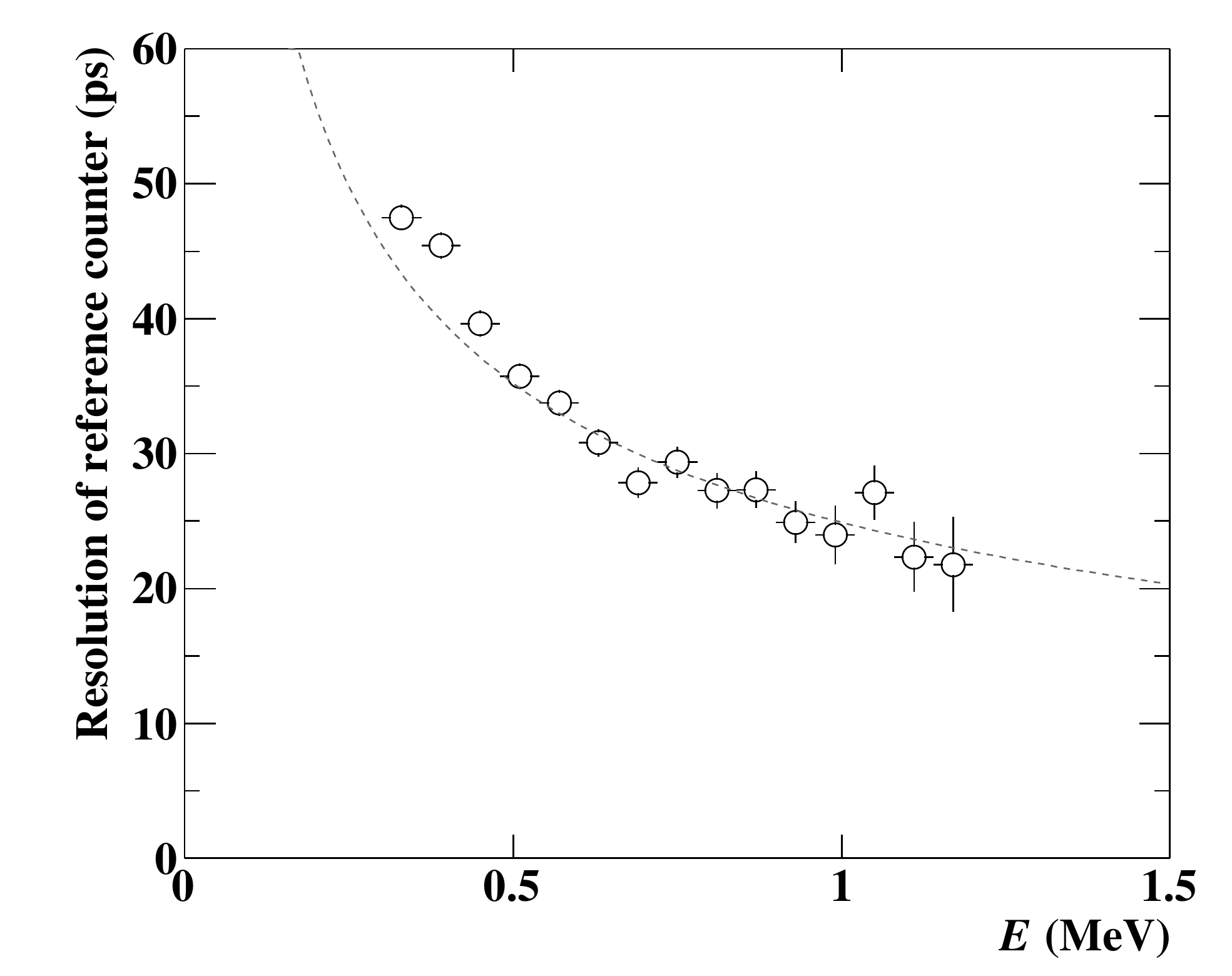}
\caption{Time resolution of the reference counter as a function of the deposited energy. The dashed line is $\sigma_\mathrm{ref}(E) = \sigma_{\mathrm{ref}}^\mathrm{1 MeV}/\sqrt{E/(1~\mathrm{MeV})}$ fitted to the data for $E> 0.5$~MeV with the best-fit value of $\sigma^{\mathrm{1 MeV}}_\mathrm{ref}=24.9\pm0.3$~ps.}
\label{fig:rc}
\end{figure}

The time resolution of the reference counter is measured by means of another small counter consisting of $25\times12\times5~\mathrm{mm}^3$ EJ232 wrapped in Teflon tape, readout with four HPK SiPMs (S10362-33-050C), two connected in series at each end\footnote{This counter was built by A.~Stoykov \cite{stoykov_NDIP}.}.  The $\Delta t$ distribution with this counter has $\sigma_{\Delta t} = 41.5\pm 0.5$~ps with our standard cut on the energy deposited in the reference counter ($>0.5$~MeV). The time resolution of this counter is evaluated to be $30.6\pm 0.4 \mathrm{(stat.)}\pm 1.0\mathrm{(sys.)}$~ps by measuring the distribution of time difference between the two ends ($(t_1-t_2)/2$) with a correction for the finite spread of the electron impact point. Then, the resolution of the reference counter is extracted, $\sigma_\mathrm{ref} = 28.0\pm 1.4$~ps. 
The dependence of the time resolution in the reference counter on the deposited energy is shown in Fig.~\ref{fig:rc}.
%

The time resolution of the test counter is cross-checked by means of the $(t_1-t_2)/2$ distribution. 
The resolutions evaluated with the two methods are consistent each other within their uncertainties.

The time resolution measured at the center of the counter is used for the comparison among different counter configurations, while the position dependence of the time resolution is described in Sec.~\ref{positiondependence}.

\subsubsection{Scintillator comparison}
A comparative study of scintillator materials is carried out with $60\times 30\times 5~\mathrm{mm^3}$ sized counters with the old-type HPK SiPMs (S10362-33-050C).
The results are summarized in Tab.~\ref{tab:scint}.
A better time resolution is obtained with BC422, which has a short rise time but a short attenuation length. 
On the other hand, a much worse resolution is obtained with the fastest scintillator, BC422Q.

The counter-size dependence of the time resolution was also measured. 
BC422 gives the highest resolution for each counter size up to $120\times 40\times 5~\mathrm{mm^3}$. 
In the following studies, BC422 is used.

\begin{table}[!t]
\renewcommand{\arraystretch}{1.3}
\caption{Scintillator comparison.}
\label{tab:scint}
\centering
  \begin{minipage}{1\linewidth}
   \renewcommand{\thefootnote}{\alph{footnote})}	
   \renewcommand{\thempfootnote}{\alph{mpfootnote})}	
\centering
\begin{tabular}{@{}lcccc}
\hline
\textbf{Properties}  & \textbf{BC418} & \textbf{BC420} & \textbf{BC422} & \textbf{BC422Q} \\
\hline \hline
 Light Output\footnotemark[1] (\% Anthracene) & $67$ & $64$ & $55$ & $19$\\
 Rise Time\footnotemark[1]\footnotemark[2] (ns) & $0.5$ & $0.5$ & $0.35$ & $0.11$\\
 Decay Time\footnotemark[1] (ns) & $1.4$ & $1.5$ & $1.6$ & $0.7$\\
 Peak Wavelength\footnotemark[1] (nm) & $391$ & $391$ & $370$ & $370$\\
 Attenuation Length\footnotemark[1] (cm) & $100$ & $110$ & $8$ & $8$\\
 Time Resolution\footnotemark[3] (ps) & $48\pm2$ & $51\pm2$ & $43\pm2$ & $66\pm3$\\
\hline
\end{tabular}
  \vspace{-0.2cm}
  \footnotetext[1]{From Saint-Gobain catalog.}
  \footnotetext[2]{Those values are dominated by the measurement setup. The intrinsic values are much faster. For example, a BC422 rise time of $<20$~ps was reported in \cite{BC422RiseTime}.}
  \footnotetext[3]{Measured value with $60\times 30\times 5~\mathrm{mm^3}$ sized counter. The uncertainties include the common systematic uncertainty of 1~ps due to the reference counter.}
\end{minipage}
\end{table}

\subsubsection{Counter size dependence}
The time resolutions were measured for different lengths and widths of counters, from $60\times 30\times 5$ to $120\times 40\times 5~\mathrm{mm^3}$ (longest) and to $90\times 50\times5~\mathrm{mm^3}$ (widest). 
The thickness was fixed at 5~mm since the time resolution is nearly insensitive to the scintillator thickness; for a given size of SiPM, the smaller energy deposition in a thinner counter is compensated by the larger fractional sensor coverage, defined by the ratio of the total SiPM active area to the cross section of the scintillator.
In this study, the old-type HPK SiPMs (S10362-33-050C) were used.

\begin{figure}[!t]
\centering
\includegraphics[width=3.4in]{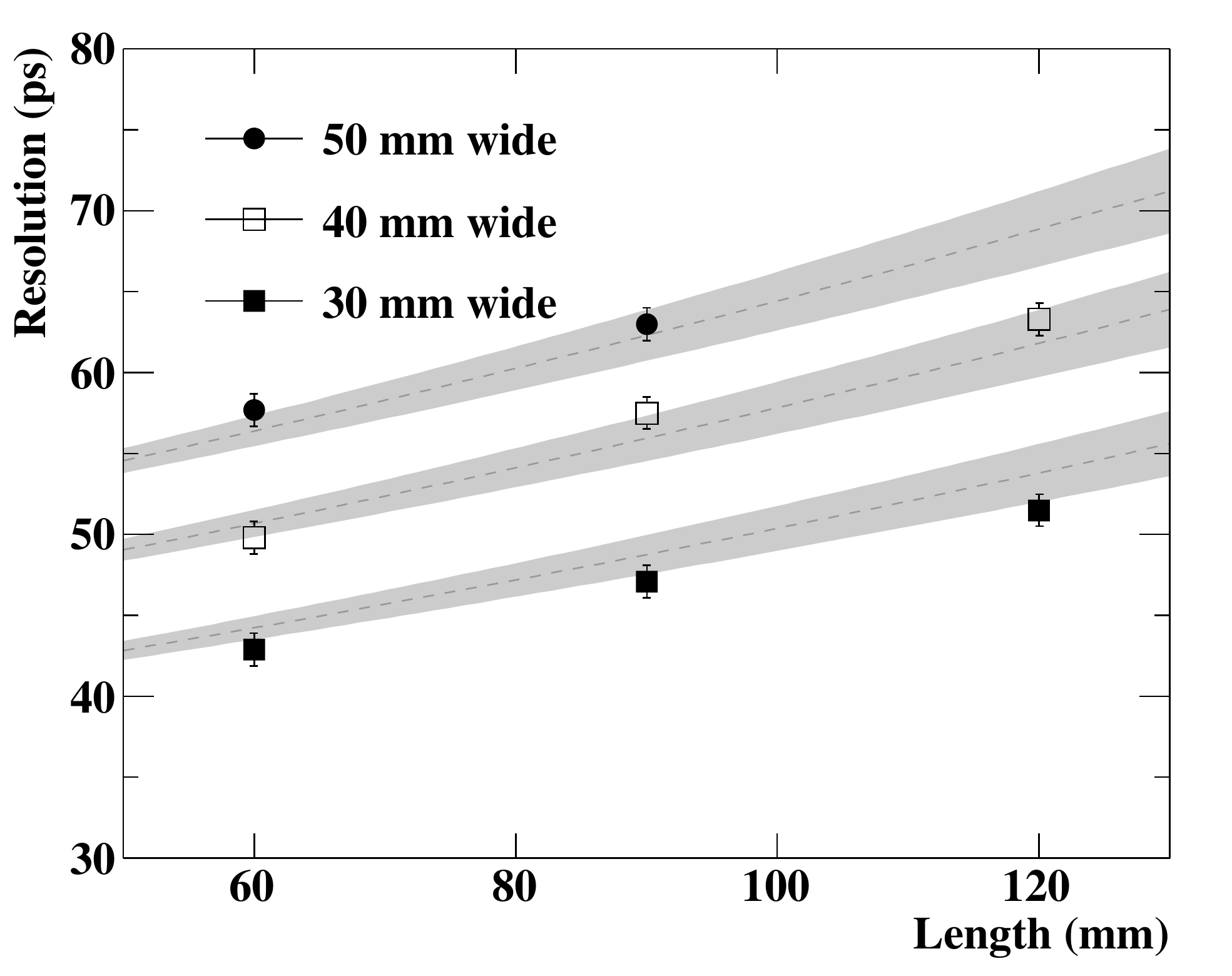}
\caption{Dependence of counter time resolution on the size.
The superimposed curves are the dependence expected from the detected photon statistics with the best-fit values of the intrinsic resolution and the correction factor for the effective path length. The shaded band shows the uncertainty. See text for the detailed description of the model.}
\label{fig:sizeDependence}
\end{figure}

The counter-size dependence is shown in Fig.~\ref{fig:sizeDependence}.
A good time resolution
is obtained even for the longest (120~mm) counters, while relatively high dependence on the counter width is observed.
The length dependence dominantly reflects the light attenuation during the propagation in the scintillator.
On the other hand, the width dependence reflects the effect of sensor coverage.
Assuming that the time resolution is proportional to the inverse of the square root of the number of detected photons, the dependence is expected to be given as
 \begin{IEEEeqnarray}{lCr}
 \sigma(k, L) =   	
	\sqrt{\frac{\sigma_\mathrm{full,0}^2 }{k \cdot \mathrm{e}^{-L\cdot f/2 	
	\lambda_\mathrm{att}}} + \sigma_\mathrm{elec}^2},
\label{sigmadep}
\end{IEEEeqnarray}
where $k$ is the fractional sensor coverage, $\lambda_\mathrm{att} = 8$~cm is the attenuation length of BC422, and $f$ is a correction factor of the counter length for the effective path length of photons propagating in the scintillator.
The value $\sigma_\mathrm{full,0}$ is the intrinsic time resolution extrapolated to the case with full coverage and no attenuation, and $\sigma_\mathrm{elec} = 11$~ps is the measured contribution from the electronics jitter.
While the upper limit of $f$ is estimated to be $< 1.3$ from the average emission angle of photons reaching the end of scintillator with total reflection and also from the measured effective velocity of the scintillation light (see \ref{positiondependence}), it is difficult to evaluate it analytically. If the formula is fit to the measured dependence with $\sigma_\mathrm{full,0}$ and $f$ as floating parameters, the best-fit values are given as 
$\sigma_\mathrm{full,0}=14.8\pm0.4$~ps and $f=1.10 \pm 0.09$. The best-fit curves are superimposed in Fig.~\ref{fig:sizeDependence}.
The size dependence of the time resolution is well understood by the photon statistics.
The average number of primary fired cells\footnote{This is equivalent to the number of photo-electrons in the case of PMTs.} is measured to be $49\pm3$ per SiPM for the $60\times 30\times 5$-sized counter at an over-voltage of $2$~V.

\begin{figure}[!t]
\centering
\includegraphics[width=3.4in]{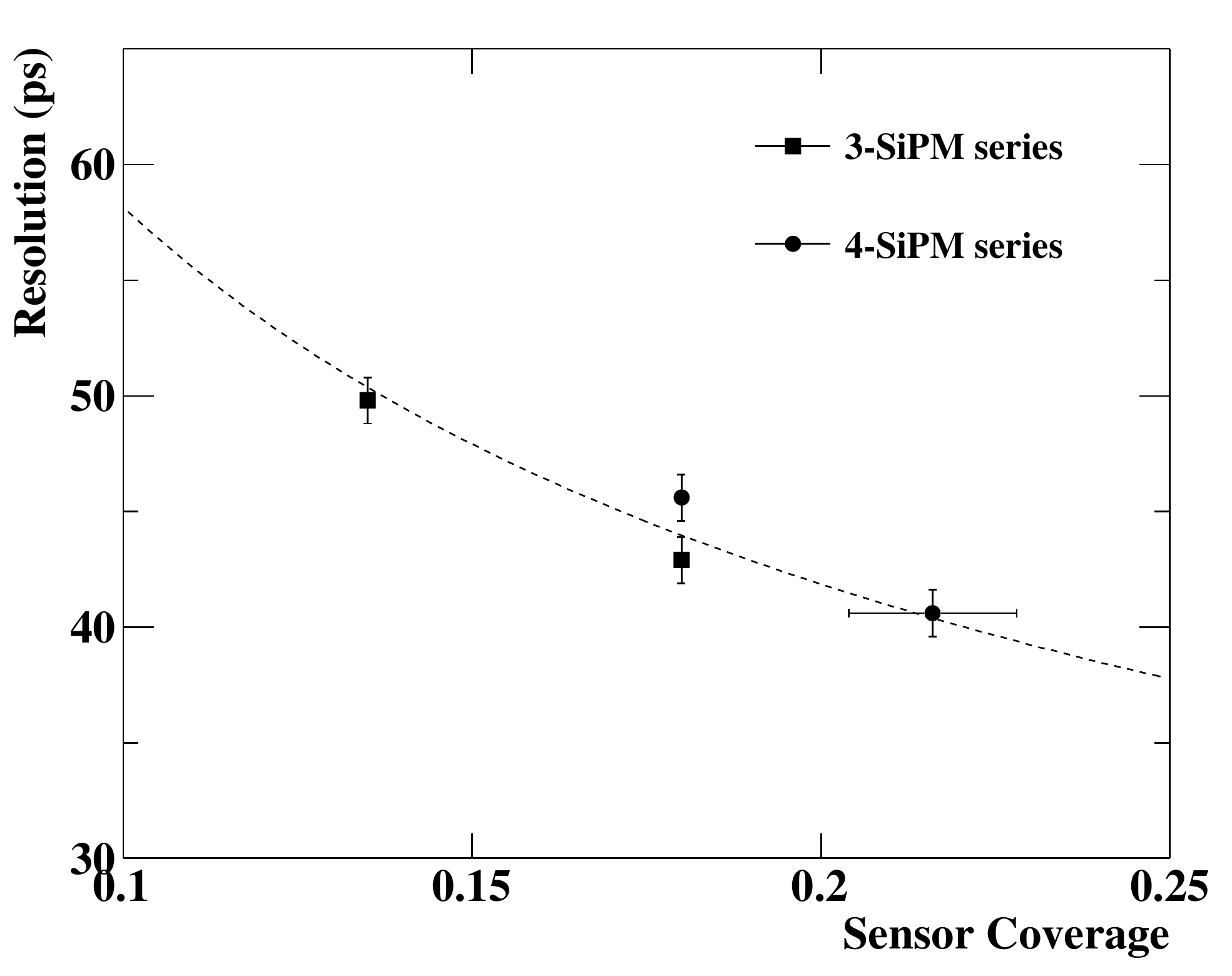}
\caption{Time resolutions for 60-mm long counters as a function of the sensor coverage. Two sizes of counters, 30 and 40 mm wide, and two configurations of SiPM connection, 3 and 4 SiPMs with series connection, were tested. The surface-mount type HPK SiPMs (S10931-050P) were used for the 30-mm wide 4-SiPM counter because of space limitation. The lower PDE (about 10\% lower than that of S10362-33-050C) is taken into account as a reduction of the coverage. The dashed curve shows the best-fit function of 
$\sigma(k) = \sqrt{\sigma_\mathrm{full}^2/k + \sigma_\mathrm{elec}^2}$.
}
\label{fig:coverageDependence}
\end{figure}

To investigate further the dependence of the time resolution on the photon statistics, we tested counter configurations with four SiPMs connected in series (in total, eight SiPMs attached to a counter). 
The resolutions of the 4-SiPM configuration as well as those of different counter sizes are plotted in Fig.~\ref{fig:coverageDependence} as a function of the sensor coverage.
The dependence follows closely $\sigma(k) = \sqrt{\sigma_\mathrm{full}^2/k + \sigma_\mathrm{elec}^2}$.
The best-fit value $\sigma_\mathrm{full}=18.1\pm0.2$~ps indicates that the intrinsic time resolution of 6-cm long counters for full coverage of the scintillator ends.


\subsubsection{Bias dependence and SiPM comparison}
\begin{figure}[!t]
\centering
\includegraphics[width=3.4in]{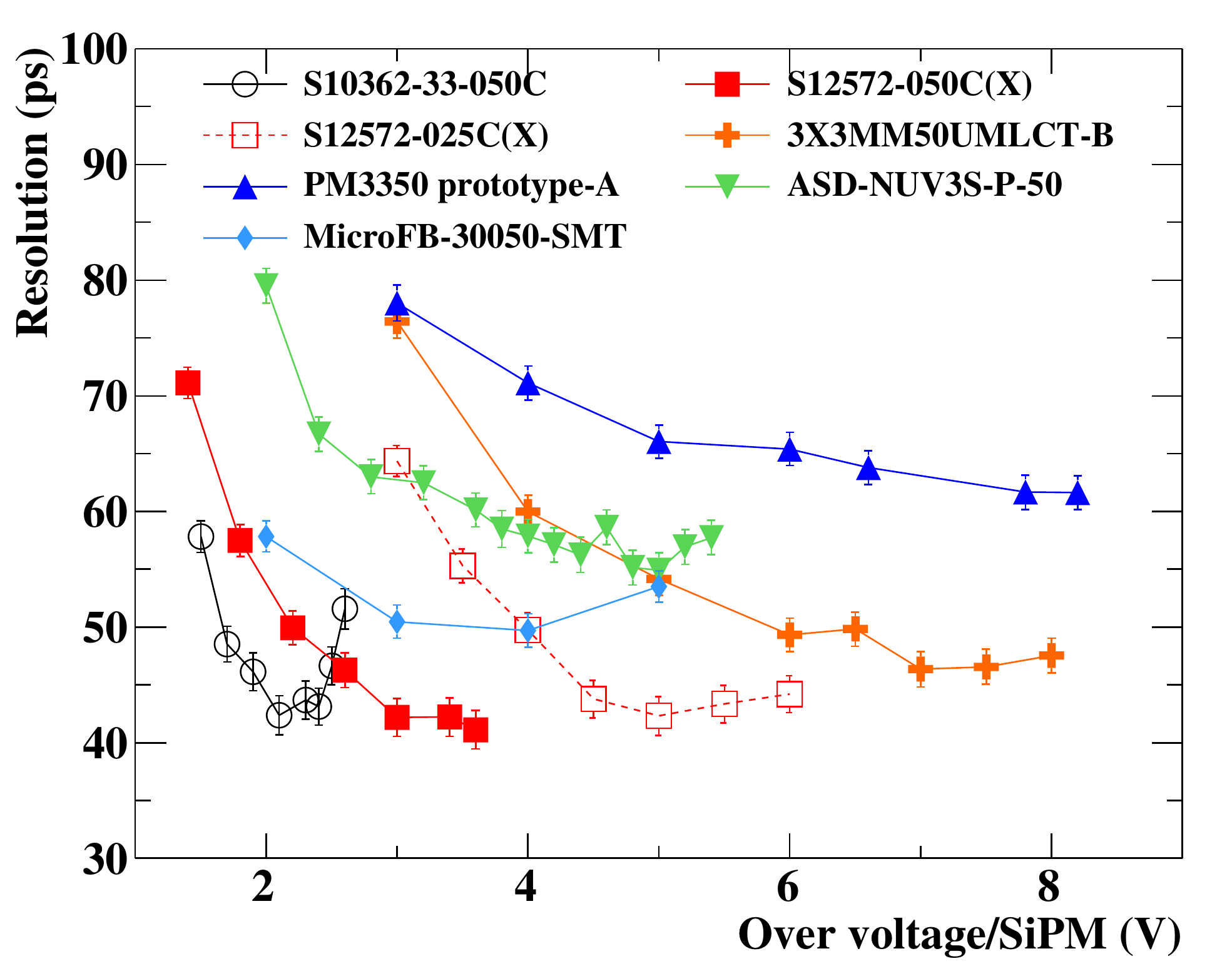}
\caption{Counter time resolution for different types of SiPMs as a function of over-voltage applied to each SiPM in a chain. 
}
\label{fig:resolution}
\end{figure}

Fig.~\ref{fig:resolution} shows the time resolutions of $60\times 30\times 5~\mathrm{mm^3}$ sized counters measured with different types of SiPMs  as a function of over-voltage per SiPM in the series readout. 
As a general trend, the resolutions improve with over-voltage.
This is due to a combination of the following factors: the improvement in SPTR due to the faster development of the avalanche \cite{collazuol_nima,puill_nima,avalancheModel}, the increase in PDE, and the improvement in the signal-to-noise ratio due to the higher gain.
The time resolution saturates with the saturation of those effects as shown in Fig.~\ref{fig:pde}. 

With the old-type HPK SiPMs, the applicable over-voltage is limited to $\sim\!2.5$~V; above this voltage, the current blows up due to the increased after-pulsing and the time resolution is degraded.
For the new types of HPK SiPMs, the time-resolution improvement continues up to higher over-voltages. 
However, the best resolution is comparable with that of the old type.
Similar behavior is observed for the other manufacturer\rq{s} SiPMs, although 
the best time resolutions are significantly worse than those of HPK\rq{s}.
The best resolution ($\sigma=42\pm2$~ps) is obtained with the new-type standard HPK SiPM.

\subsection{Rise time and temperature stability}

\begin{figure}[!t]
\centering
\subfloat[Rise time]{\includegraphics[width=3.4in]{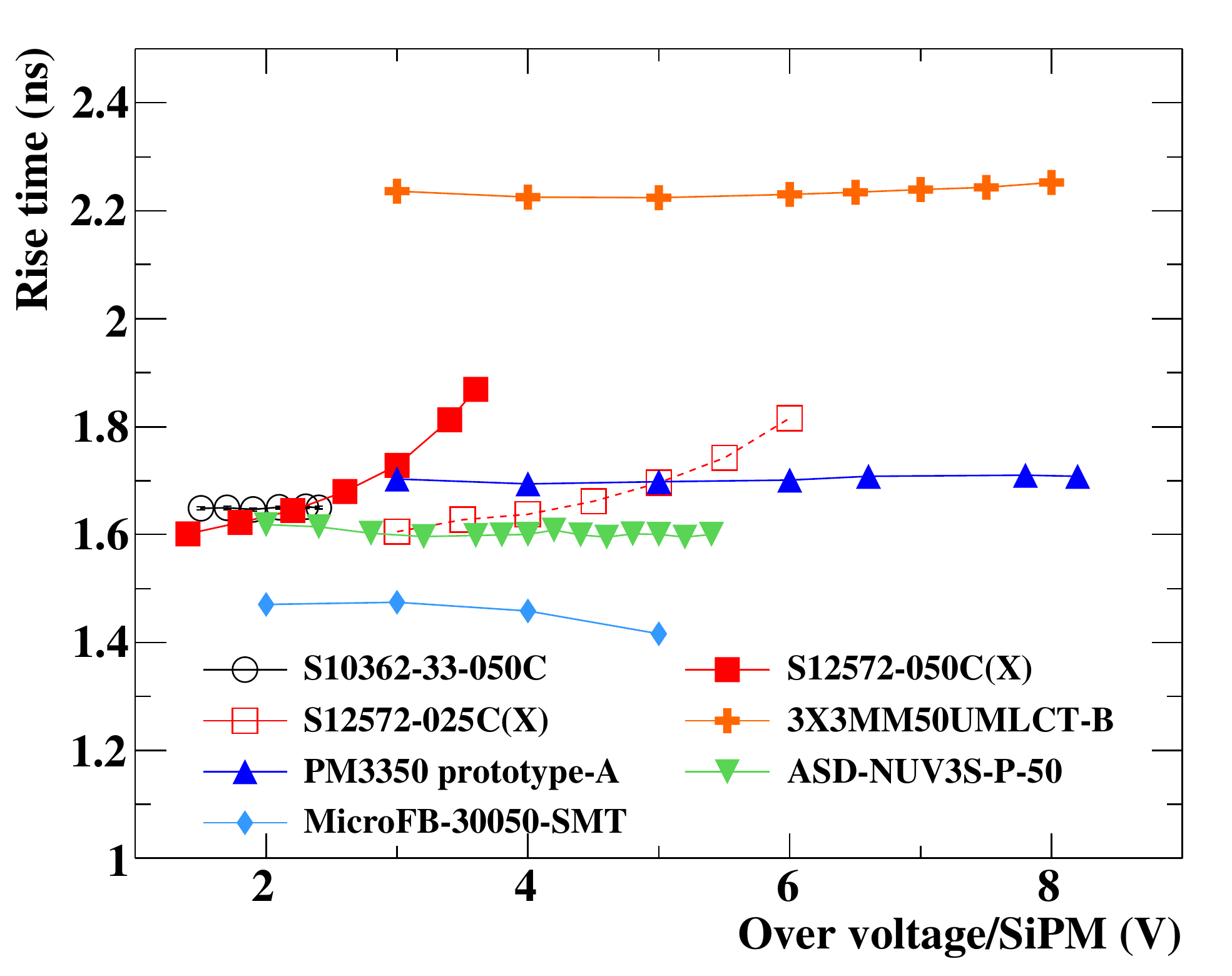}%
\label{fig:risetime}}
\hfil
\subfloat[Time center]{\includegraphics[width=3.4in]{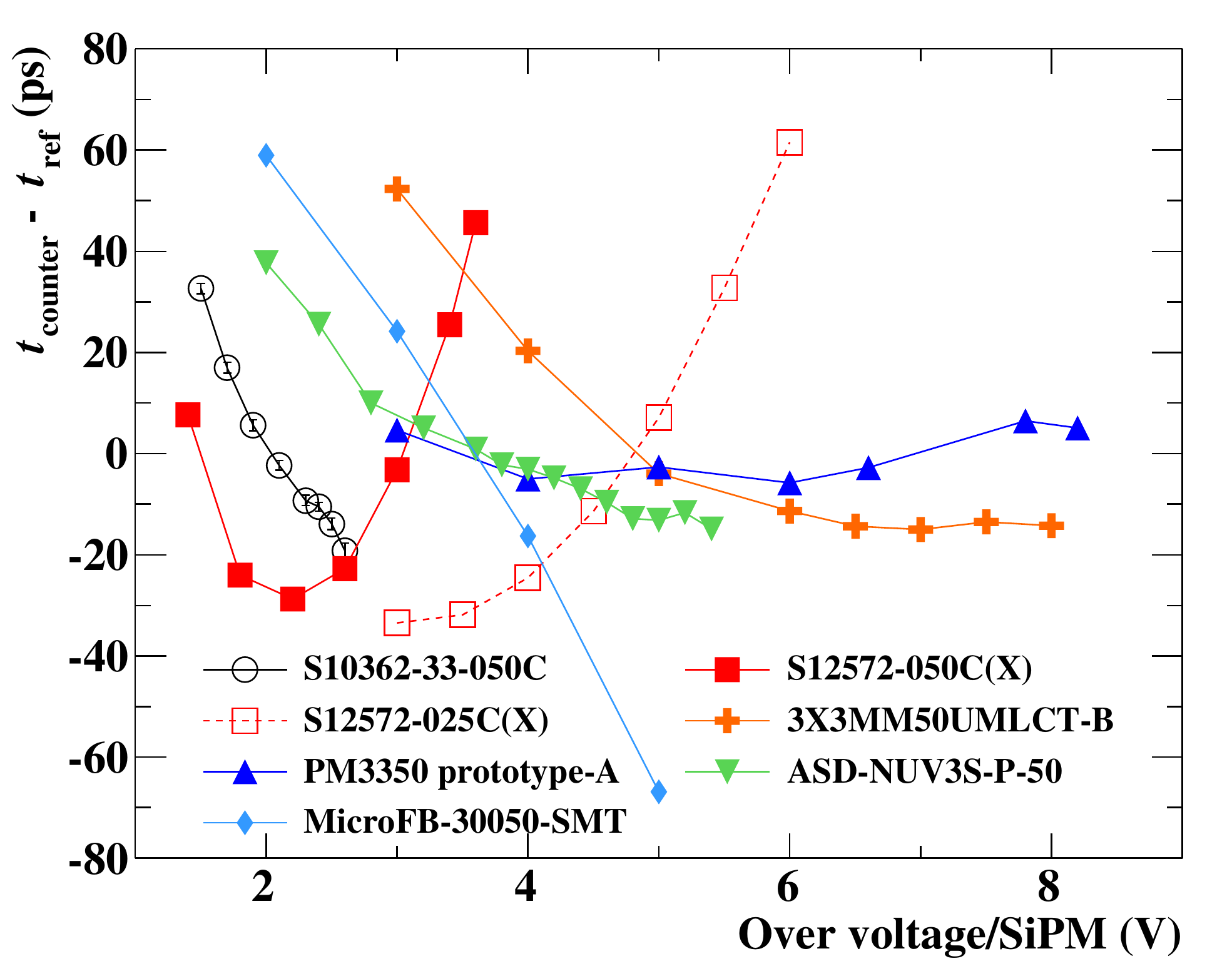}%
\label{fig:timecenter}}
\caption{Rise time of the detector signal and center value of the $\Delta t$ distribution for different types of SiPMs as a function of over-voltage.}
\label{fig:timeStability}
\end{figure}

Fig.~\ref{fig:risetime} shows the dependence of the rise time (10\% to 90\%) on the over-voltage.
The measured rise times are mainly limited by the RC time constant, from the total capacitance of the SiPM-chain plus the cable capacitance and the input impedance of the amplifier, convolved with the finite time width of the scintillation process. 
The intrinsic SiPM rise time should decrease with increasing over-voltage because of the faster build-up of the avalanche process \cite{avalancheDiffuse,avalanchePhoto,avalancheModel}.
The observed over-voltage dependence of the SensL SiPMs, which give the fastest rise time due to the fast output scheme, suggests this effect.
While the main contribution from the readout circuit comes from the long coaxial cable ($\sim\!7$\% increase in the rise time), its impact on the time resolution is measured to be less than 2\%.

We observe softening of the leading edge at high over-voltages for the new-type HPK SiPM.
The pulse shape variation results in a larger over-voltage dependence of the signal arrival time.
Fig.~\ref{fig:timecenter} shows the over-voltage dependence of the measured signal arrival time with respect to the time measured by the reference counter.
Generally, as the bias voltage increases, the signal arrives earlier for the same reason that the rise time decreases. 
However, due to the softening of the pulse shape, the signal arrival time with the new-type HPK SiPM is delayed again at higher over-voltages.

It is well known that the breakdown voltage of SiPMs depends on temperature.
Therefore, the over-voltage dependence of the signal arrival time would also result in a temperature dependence of this parameter if the bias voltage is fixed.
We measured the temperature dependence of signal arrival time by directly varying the temperature in the range of 20$^\circ$--40$^\circ$C and compared the temperature dependence with the over-voltage dependence.
The results suggest that the two effects are equivalent; the temperature dependence of the signal arrival time only stems from the change of over-voltage due to the change of the breakdown voltage; other effects, such as pulse shape variation due to change of the quench resistance, are negligible in this temperature range.
  
Table~\ref{tab:tempcoef} summarizes the temperature coefficients of the signal-arrival-time drift at each optimal resolution point.
The deterioration of the temperature coefficient for the new standard-type HPK SiPMs
could be an issue in some practical applications unless active control of bias voltages according to the temperature is possible. 
In contrast, the trench-type HPK SiPMs show a much better stability. 
Since SiPMs from other manufacturers have 
small temperature coefficients of the breakdown voltage, the fluctuation of the signal arrival time due to temperature variation is negligible compared with their intrinsic time resolutions.

\begin{table}[!t]
\renewcommand{\arraystretch}{1.3}
\caption{Temperature coefficients of signal arrival time drift (ps/$^\circ$C).}
\label{tab:tempcoef}
\centering
\begin{tabular}{@{}cc@{~~}c@{~~}cccc@{}}
\hline
\vspace{-2mm}\\
\shortstack{HPK\\Old} & \shortstack{HPK\\New} & \shortstack{HPK\\New-25\,$\upmu$m} &\shortstack{HPK\\Trench} & KETEK & \shortstack{AdvanSiD\\NUV} & \shortstack{SensL\\B-series}\\
\hline \hline
2.5 & 5.5 & 2.8 & 0.1 & 0.1 & 0.2 & 0.8 \\
\hline
\end{tabular}
\end{table}

\subsection{Position dependence}
\label{positiondependence}
\begin{figure}[!t]
\centering
\subfloat[Time resolution]{\includegraphics[width=3.5in]{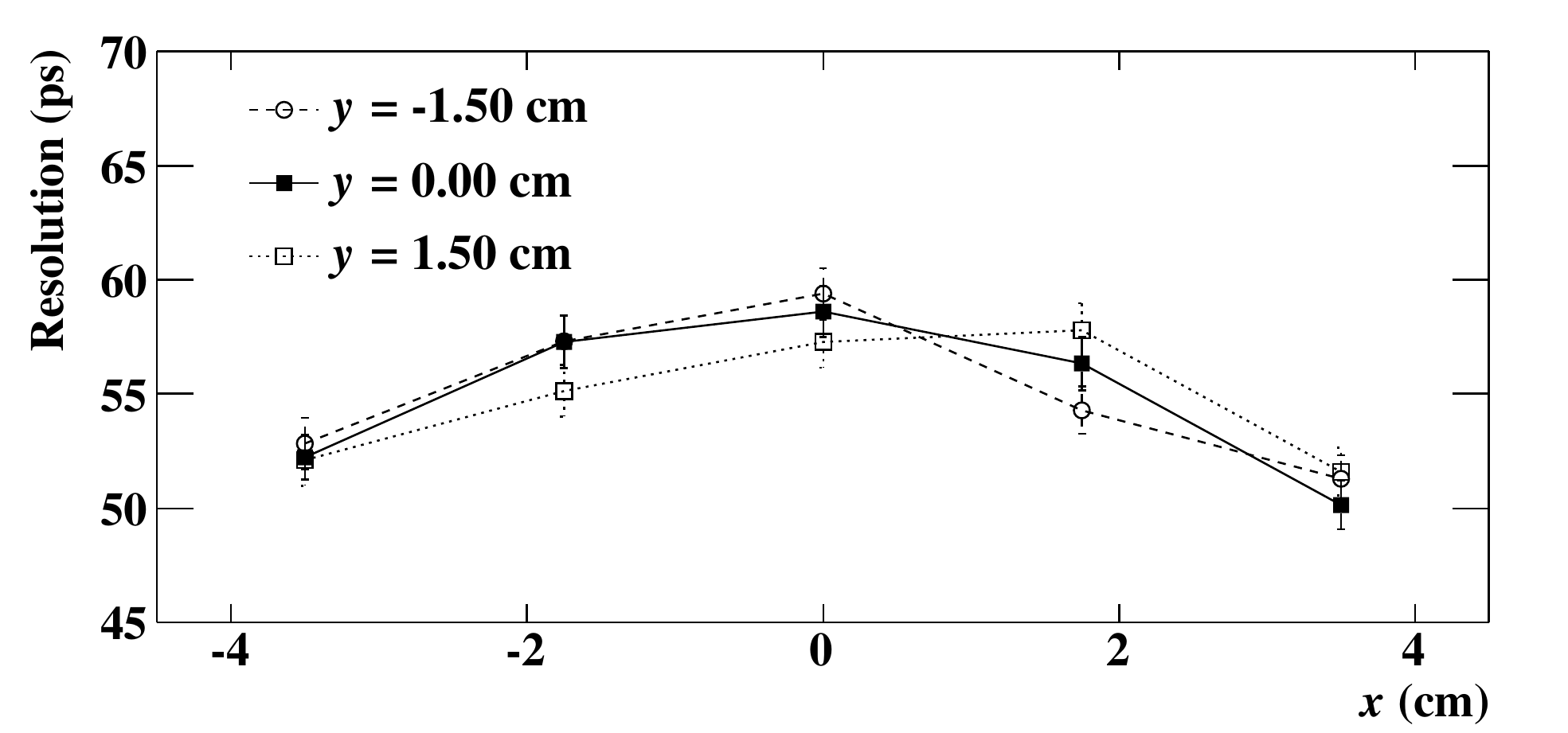}%
\label{fig:positionRes}}
\hfil
\subfloat[Time center]{\includegraphics[width=3.5in]{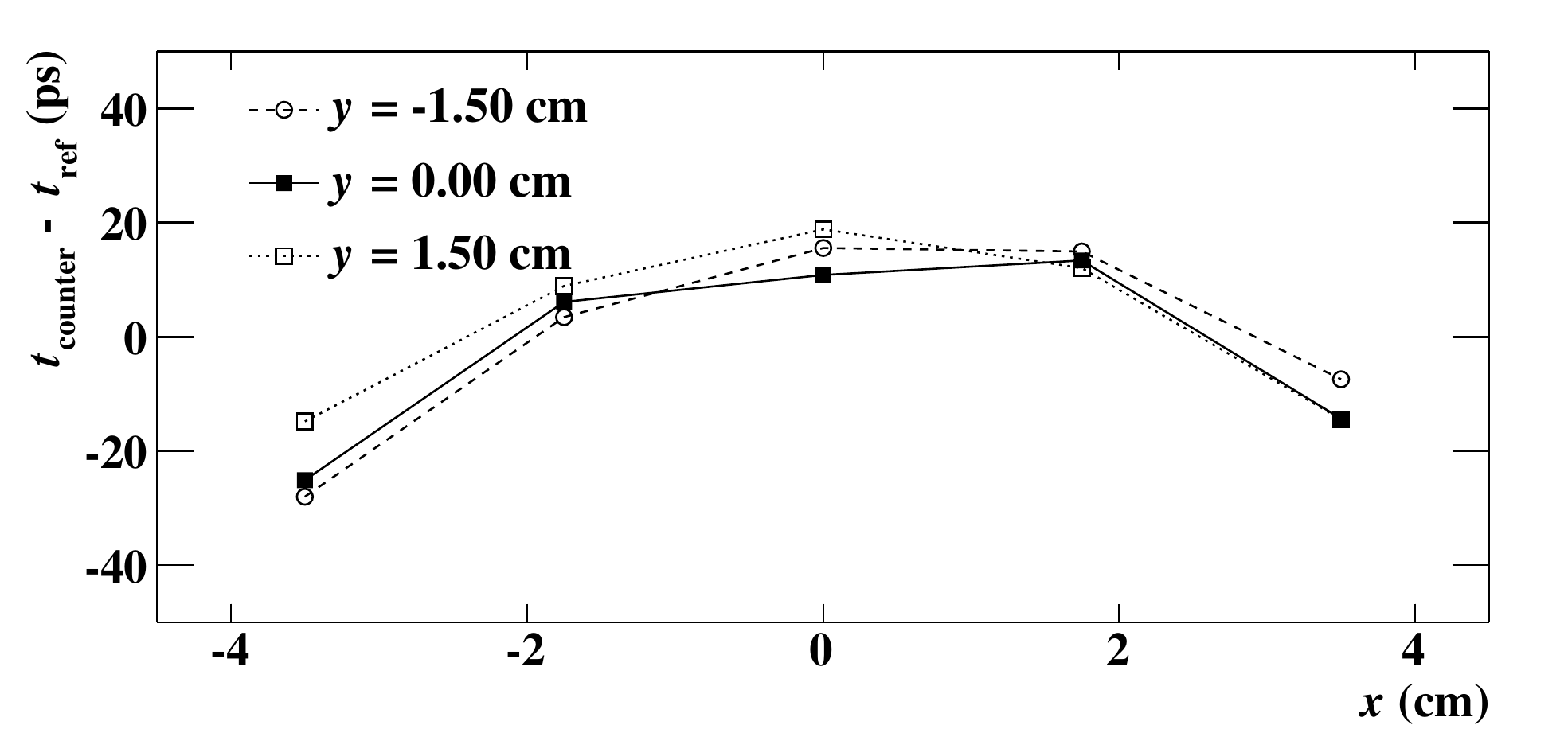}%
\label{fig:positionCenter}}
\caption{Position dependence of time measurement for $90\times40\times5~\mathrm{mm^3}$ counter. $x$ and $y$ are the coordinates of the center position of the electron irradiation in length and width directions, respectively.}
\label{fig:positionDep}
\end{figure}

The position dependence of the counter response was investigated by repeating the measurements at 15 different positions.
Fig.~\ref{fig:positionDep} shows the result for the $90\times 40\times 5~\mathrm{mm^3}$ counter.

The resolutions are better near the ends of the scintillator where the SiPMs are attached. 
This is because of the increase of the number of direct photons from the emission point before any reflection.
On the other hand, the dependence of the resolution on the position along the width direction does not exceed more than 10\%. 

\begin{figure}[!t]
\centering
\includegraphics[width=2.9in]{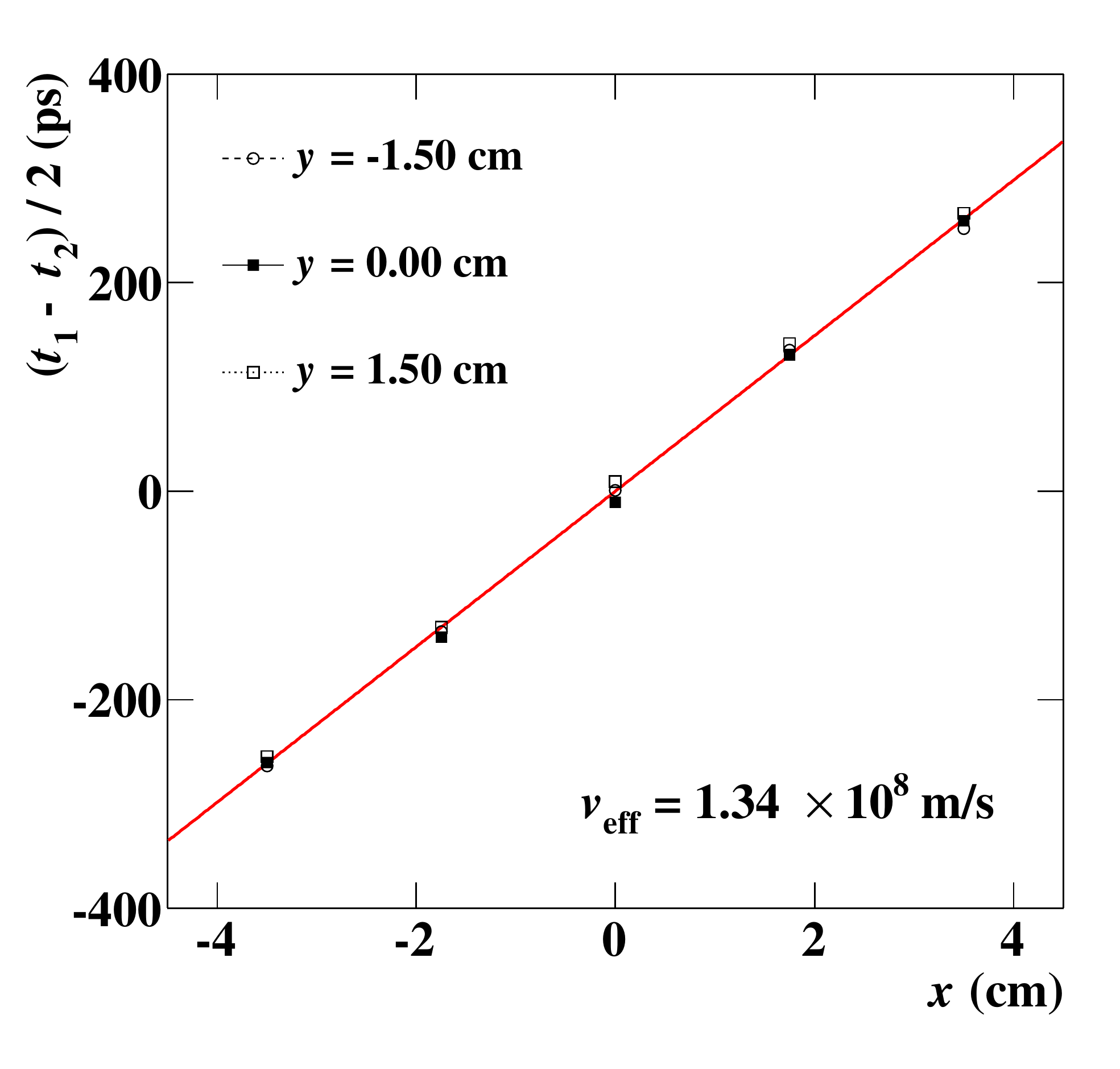}
\caption{Relation between the electron incident point and the time difference between the two signals at the end of a counter. The solid line shows the best linear fit to the data. The effective speed of light $v_\mathrm{eff}$ in the scintillator is computed from the slope of the linear function.}
\label{fig:positionRec}
\end{figure}

A variation of $\sim\!50$~ps in the reconstructed signal times is observed along the length of the scintillator.
However, this position dependence can be corrected for by using a reconstructed impact position. As shown in Fig.~\ref{fig:positionRec}, the difference in the measured times at the two ends can be used to reconstruct the longitudinal position with a resolution of 8~mm.
The overall resolution, $\sigma_\mathrm{all}=58.1\pm0.3$~ps when no correction is applied, improves slightly to $57.4\pm0.3$~ps if the correction is applied. If the position of the impact point is known precisely, it improves to $55.9\pm 0.3$~ps.


\section{Discussion}

\subsection{Limitation in time resolution}
\label{discussiontr}
The dependence of the time resolution on the sensor coverage indicates that the time resolution is proportional to the inverse of the square root of the total number of photons detected by the SiPMs (namely, the number of primary fired pixels).
The counter-size dependence can be understood from the photon-counting statistics.
The higher resolution of HPK SiPMs is also understood from the same argument because they have higher PDEs than others.
Therefore, for the configuration of our counter, the time resolution is limited by photon counting statistics.
This is, for example, the same conclusion reached in \cite{sipmModel}, where a comprehensive model to predict the time resolution of small SiPM-based scintillation counters was proposed.

The number of detected photons is expected to be proportional to the sensor coverage and the energy deposited in the scintillator ($k\cdot E$).
Therefore, the relation between the counter time resolution, the sensor coverage and the deposited energy should follow the relationship given by 
$\sigma_{\mathrm{full}}^{\mathrm{1 MeV}}=\sigma\sqrt{k(E/\mathrm{MeV})}~\mathrm{ps}
$, which can be used to compare counter intrinsic resolutions. For many applications of PMT-based counters and small-sized SiPM-based counters such as in \cite{muSR}, the sensor fractional coverage is equal to one, while for our case it is significantly smaller ($k=0.18$ for a 3-cm wide, 3-SiPM configuration counter). 
From the best obtained resolution and the mean energy deposition of $0.95$~MeV, our result corresponds to $\sigma_{\mathrm{full}}^{\mathrm{1 MeV}} = 18~\mathrm{ps}\,(\mathrm{MeV}/E)^{1/2}$.
This is the best resolution ever achieved, as discussed in \cite{muSR}.
Considering the effect of scintillation absorption in the larger scintillators, 
the intrinsic performance of the scintillation counter is improved.
The measured counter-size dependence suggests that the intrinsic resolution would reach 
$\sigma_{\mathrm{full}}^{\mathrm{1 MeV}} = 14.8~\mathrm{ps}\,(\mathrm{MeV}/E)^{1/2}$ for small counters for which the effect of the light attenuation can be neglected.
This is in fact consistent with the value measured with the reference counter ($k=0.36$) shown in Fig.~\ref{fig:rc}.
The main contributions to the systematic uncertainty come from the reference counter resolution (3\% in $\sigma_{\mathrm{full}}^{\mathrm{1 MeV}}$), the energy scale (2\%) and the extrapolation (3\%). The total systematic uncertainty on $\sigma_{\mathrm{full}}^{\mathrm{1 MeV}}$ is estimated to be 4\%.

In addition to the limitation from photon-counting statistics, we observed other factors contributing to the time resolution depending on the bias voltage. 
The dark noise (thermal noise and after-pulses) induces a fluctuation of the baseline.
For the time measurement, high-pass filters such as the pole-zero cancellation, efficiently suppress the effect.
However, a degradation of the time resolution starts when the dark current exceeds $\sim\! 10$~$\upmu$A.
After-pulsing would also be an issue in high count rate experiments.

The rise time increase and the significant drift of signal time observed with the new-standard-type HPK SiPMs can be understood as a consequence of the high-probability of the cross-talk process.
These observations suggest that the stochastic process of the cross-talk causes an event-by-event fluctuation of the signal pulse shape.
We consider this to be a reason for the limited improvement of the new-type HPK SiPM time resolution in spite of the improved PDE.
The situation with a very high cross-talk probability was not considered in the model of \cite{sipmModel}, while the recently developed technique for after-pulse suppression pushes the SiPM usage into such situations.

The effect of electronics, in particular the time calibration of each sampling point of DRS4, is evaluated to be $\sigma\approx 10$~ps. 
Though not a major factor, it is expected to be reduced to $O(1~\mathrm{ps})$ \cite{drstime}.

\subsection{Possible improvements}

Manufacturers of SiPMs continuously improve their devices.
HPK presented improvements in rise time and pulse amplitude for their trench-type SiPMs  \cite{hamamatsu_2013}.
The fill factor of trench-type will be as large as that for the current standard type, namely $>\! 60$\%, while that for the standard type will exceed 80\%.
Other manufacturers will also provide improved devices, in both their PDEs and SPTRs.
Those new developments will surely improve the intrinsic counter resolution, $\sigma_{\mathrm{full}}^{\mathrm{1 MeV}}$, and thus, the total time resolution.

Another important potential point of progress is in reducing cost, which will allow other improvements in practical applications.
A straightforward way to increase the photon counting statistics is to increase the number of sensors to get a larger coverage, as in Fig.\ref{fig:coverageDependence}.  
We will test configurations with more than four SiPMs at each side.
Using larger SiPMs in series connection is another approach: using $4\times4~\mathrm{mm^2}$ SiPMs 
increases the photon statistics by a factor 1.78, resulting in an expected improvement in time resolution of about 33\%.

\subsection{Application to high energy experiments}
The counters we have developed can be applied to high energy physics experiments such as MEG II.
In such an experiment, counters can be placed stacked or staggered along the path of particle trajectories, relying on the thinness of the counters to reduce scattering \cite{metcjinst2014}. 
By measuring $N_{hit}$ hit times of a particle with multiple counters and combining these measurements, the time resolution can be improved according
to $1/\sqrt{N_{hit}}$; in MEG II $\overline{N_{hit}}>4$ is expected.
This improvement works also for other contributions such as the time calibration among counters and the synchronization among the readout electronics channels, which are expected to be important contributions to the final resolution in many practical cases (such as the timing counter system in MEG \cite{tc2}). 
Therefore the expected performance of the timing counter system is $\lesssim 42/2 = 21$~ps, significantly better than traditional PMT-based detector using 
long bars, like MEG \cite{tc} and MICE \cite{mice}, which achieve resolutions $\sim 50$ ps. 

This improvement comes at the cost of an increase in the number of readout channels.
The total number of readout channels in MEG II is expected to be 1\,024 (512 counters), more than an order of magnitude increase from MEG (60 channels).

\section{Conclusions}

The time resolutions of PMT-based scintillation counters are predominantly limited by photon-counting statistics; it is true also for SiPM-based counters.

The small size of SiPMs, which is the main drawback of SiPM readout in large-sized detectors, can be compensated by using several SiPMs connected in series.  
Series connection is superior to parallel connection for time measurement due to the reduction in capacitance and the automatic equalization of over-voltages among sensors. 

Signal processing with pole-zero-cancellation analog shaping followed by a high-speed-sampling waveform digitizer is highly optimized for time measurement
with SiPM readout.

From the comparative studies of scintillator materials and NUV-sensitive SiPMs,
the best time resolution is obtained with the combination of the fast scintillator BC422 and the new-type HPK SiPMs with the highest PDE in the NUV region.
An excellent resolution of $\sigma =42 \pm 2$~ps at 1~MeV energy deposition is obtained for a counter with dimensions $60\times 30 \times 5~\mathrm{mm^3}$ and three SiPMs at each end of the scintillator.
Further improvement is possible by increasing the number of SiPMs attached to the scintillator.

The counters we have developed, with very good time resolution and dimensions from $60\times 30 \times 5$ to $120\times 40 \times 5~\mathrm{mm^3}$ can be applied to high energy physics experiments such as the MEG II experiment. 
In such an experiment, by measuring each particle\rq{}s time with multiple counters, the final time resolution can be significantly improved with respect to that of a single counter, being superior to those of conventional PMT-based detectors.

\section*{Acknowledgment}
The authors would like to thank the solid state division of Hamamatsu Photonics for providing us test samples and Alexey Stoykov for his help at the early stage of the R\&D. 
We also wish to thank the Electronics \& Measuring Systems Group and the Detector Group in the Laboratory of Particle Physics at the Paul Scherrer Institut and 
the mechanical and electronics workshops at the INFN Section of Genova for their valuable help.

\ifCLASSOPTIONcaptionsoff
  \newpage
\fi



\bibliographystyle{IEEEtran}
\bibliography{IEEEabrv,IEEEtran-2014.bib}

\end{document}